\documentclass[aps,prc,twocolumn,showpacs,floatfix,superscriptaddress,amsmath,amssymb,nofootinbib,longbibliography]{revtex4-1}
   
\usepackage{amsmath}
\usepackage{hyperref}
\usepackage{color}
\usepackage{graphicx}

\newcommand{\be}{\begin{equation}}
\newcommand{\ee}{  \end{equation}}
\newcommand{\ba}{\begin{eqnarray}}
\newcommand{\ea}{  \end{eqnarray}}
\newcommand{\bas}{\begin{eqnarray*}}
\newcommand{\eas}{  \end{eqnarray*}}

\begin{document}

\title{Effective field theory of pairing rotations}

\author{T. Papenbrock} 

\affiliation{Department of Physics and Astronomy, University of Tennessee, Knoxville, Tennessee 37996, USA}
\affiliation{Physics Division, Oak Ridge National Laboratory, Oak Ridge, Tennessee 37831, USA}

\begin{abstract}
Pairing rotations are the low-energy excitations of finite superfluid
systems, connecting systems that differ in their number of Cooper
pairs. This paper presents a model-independent derivation of pairing
rotations within an effective theory that exploits the emergent
breaking of $U(1)$ phase symmetries. The symmetries are realized
nonlinearly and the Nambu-Goldstone modes depend only on time because
the system is finite. Semi-magic nuclei exhibit pairing rotational
bands while the pairing spectrum becomes an elliptical paraboloid for
open-shell nuclei. Model-independent relations between double
charge-exchange reactions and $\alpha$ particle capture or knockout in
open-shell nuclei are in analogy to the pair transfer reactions in a
single superfluid. Odd semi-magic nuclei are described by coupling a
fermion to the superfluid.  The leading-order theories reproduce data
for pairing rotational bands within uncertainty estimates.
\end{abstract}

\maketitle
\section{Introduction}

Atomic nuclei are finite superconductors. Hallmarks of nuclear pairing
are excitation gaps in even-even nuclei~\cite{bohr1958}, reduced
moments of inertia due to superfluidity~\cite{migdal1959}, odd-even
staggerings in many observables, and pairing vibrations~\cite{bes1966}
and rotations~\cite{nogami1964,bohr1969,bes1970,broglia1973} (see
Ref.~\cite{brink_broglia_2005} for an overview).  Pairing rotational
spectra are the analogue to rotational bands in deformed nuclei; they
are quadratic in the difference of Cooper pairs and are associated
with the Nambu-Goldstone mode of a broken $U(1)$ phase
symmetry~\cite{broglia2000,hinohara2016,potel2017}. They explain why
two-nucleon transfer is enhanced between nuclei within a
pairing-rotational
band~\cite{broglia1973,oertzen2001,potel2011,potel2013,potel2013b,shimoyama2011}.
Pairing rotational modes have been studied via pairing models, see
e.g., Refs.~\cite{bes1970,beck1972,broglia1973,matsuo1986b}, and in
Hartree-Fock-Bogoliubov~\cite{hinohara2015,hinohara2018} and
relativistic mean-field computations~\cite{kouno2021}.

In this paper we revisit pairing rotations in the framework of
effective field
theory~\cite{vankolck1999,hammer2000,bedaque2002,furnstahl2007,epelbaum2009,papenbrock2011,griesshammer2012,hammer2017,hammer2020}. This
brings simplicity and model independence to an old subject. The
approach requires us to be conscious about the breakdown scale, and
the power counting allows us to estimate or
quantify~\cite{schindler2009,furnstahl2014c} uncertainties. Open-shell
nuclei are described as two interacting superfluids starting from the
most general Lagrangian compatible with the symmetry breaking. As we
will see, the model-independent approach yields relations between
double charge-exchange reactions and two-nucleon transfer, and these
can be tested experimentally.

The approach presented in this work differs from the one by
\textcite{furnstahl2007}. That work proposed an effective field theory
for dilute Fermi systems. Here we merely exploit the dynamics of
Nambu-Goldstone modes corresponding to the emergent breaking of phase
symmetries in finite systems. Then quantum field theory reduces to
quantum mechanics~\cite{gasser1988} and a fermion only appears in odd
systems.

This paper is organized as follows. Section~\ref{sec:one} present the
effective field theory for even and odd semi-magic nuclei,
respectively.  The theory for open-shell even-even nuclei is derived
in Sect.~\ref{sec:two}. The theory is confronted with data in
Sect.~\ref{sec:app}.  The summary in Sect.~\ref{sec:sum} discusses the
main results.

\section{Effective theory for a single superfluid}
\label{sec:one}

\subsection{Even semi-magic nuclei}

\subsubsection{Leading-order Hamiltonian}
Let us consider a finite superfluid of a single fermion species with
spin $1/2$ and assume that all fermions are in Cooper pairs. Examples
are even isotopes of tin or lead, or even isotones with neutron number
$N=82$. The corresponding nuclear ground states must be invariant
under $U(1)$ phase transformations which are generated by
\be
\label{gauge}
g(\alpha) = e^{i\alpha\hat{n}} \ . 
\ee
Here $\alpha$ is the phase angle and $\hat{n}$ is the operator that
yields the number of pairs.  A finite system displays emergent rather
than spontaneous symmetry
breaking~\cite{yannouleas2007}. Nevertheless, we can follow the
standard approach to spontaneous symmetry breaking via non-linear
realizations~\cite{weinberg1968,callan1969,coleman1969,brauner2010},
and the Nambu-Goldstone mode parameterizes the coset $U(1)/1\sim U(1)$
of the broken symmetry. For finite systems, however, a tremendous
simplification occurs because the Nambu-Goldstone ``field"
$\alpha=\alpha(t)$ depends only on
time~\cite{gasser1988,papenbrock2014}, and quantum field theory thus
reduces to quantum mechanics. In our case, the phase velocity
 \be
 \dot{\alpha}   \equiv  \partial_t\alpha
 \ee
is the only quantity that can enter the  Lagrangian. 

The leading-order Lagrangian then becomes
\be
\label{LOlag}
L_{\rm LO} = {a\over 2} \dot{\alpha}^2 +n_0\dot{\alpha} \ . 
\ee
Here, $a$ and $n_0$ are low-energy constants. The constant $a$ is akin
to a mass term while $n_0$ is a constant gauge potential.  A Legendre
transformation yields the Hamiltonian
\be
\label{LOham}
H_{\rm LO} = {1\over 2a} \left(p_\alpha -n_0\right)^2 \ . 
\ee
Here, 
\be
p_\alpha\equiv {\partial L\over\partial\dot{\alpha}}
\ee
is the canonical momentum. We see that $p_\alpha$ is a constant of
motion. For an interpretation of $p_\alpha$ as the number of pairs $N$
we consider phase transformations $g(\beta)g(\alpha) = g(\alpha
+\beta)$. Thus, the phase $\alpha$ changes to $\alpha+\beta$ and this
is a nonlinear realization of the phase symmetry. Applying Noether's
theorem to infinitesimal phase transformations then yields that
$p_\alpha$ is conserved and therefore must be identified with the
number of pairs.

We quantize the Hamiltonian~(\ref{LOham}) as usual via
\be
\label{quant}
p_\alpha \to \hat{p}_\alpha=-i\partial_\alpha \ , 
\ee
(and of course also identify the pair number operator as
$\hat{n}=-i\partial_\alpha$.) Thus, the Hamiltonian is
\ba
\label{hamLO}
H &=& {1\over 2a}\left(-i\partial_\alpha -n_0\right)^2 \nonumber\\
&=& {1\over 2a}\left(\hat{n} -n_0\right)^2 
 \ea

Requiring that the wave function $\psi(\alpha)$ is single-valued under
gauge transformations
$\psi(\alpha)\to\psi'(\alpha)=e^{i\lambda\alpha}\psi(\alpha)$ with
constant $\lambda$ then shows that $n_0$ must be an
integer. Eigenfunctions are
\be
\psi_n(\alpha)=\langle \alpha|n\rangle = {1\over \sqrt{2\pi}} e^{i n\alpha} \ , 
\ee 
and these describe a system of $N$ pairs. The corresponding energies 
\be
\label{spec}
\varepsilon_n = \frac{1}{2a}(n-n_0)^2 
\ee
are in a pairing rotational band. 

Let us discuss time reversal invariance. The pair-number operator
$\hat{n}=\hat{p}_\alpha$ is even under time reversal. This implies
that the phase $\alpha$ is odd.  As $\dot{\alpha}$ is even under time
reversal, higher-order contributions to the effective theory can also
contain odd powers of the phase velocity. Under time reversal, the
eigenfunction $\psi_{n}(\alpha)\to \psi_{n}(-\alpha) =
\psi_{n}^*(\alpha) = \psi_{-n}(\alpha)$.  Formally, we could admit
negative pair numbers $n$ (and negative $n_0$), and the spectrum is
invariant under this change. In this case, we would interpret $n$ as
the number of hole pairs.

Besides the number operator $\hat{n}$, the other operator of interest
is the pair-removal operator $\hat{P}$ with matrix elements
\be
\label{pair}
\langle\alpha'|\hat{P}|\alpha\rangle =  P_0 e^{-i\alpha} \delta(\alpha'-\alpha)\ . 
\ee
Here, $P_0$ is a constant that denotes the overall strength.  Clearly
$\hat{P} |n\rangle = P_0|n-1\rangle$, and $\hat{P}^\dagger |n\rangle =
P_0^*|n+1\rangle$. The effective theory then predicts that $\langle
n+1|\hat{P}^\dagger|n\rangle =P_0^*$, i.e. pair transfer within the
nuclei of a pairing rotational band is independent of the number of
pairs in a given nucleus. This hallmark of pairing rotations has been
confirmed experimentally in two-nucleon transfer reactions,
see~\cite{broglia1973,oertzen2001,potel2013}.

\subsubsection{Power counting}
\label{sec:power}
Effective theories exploit a separation of energy scales and organize
the Hamiltonian by a power counting. In our case, the
Lagrangian~(\ref{LOlag}) is postulated to be of the low-energy scale
$\xi$ we are interested in. We thus assign the scalings
\ba
a&\sim& \xi^{-1} \ , \\
\dot{\alpha} &\sim& n_0 \xi \ , 
\ea
and it is implied that the two terms of the Lagrangian~(\ref{LOlag})
combine (or cancel) to yield the low-energy scale $\xi$. Then the
Hamiltonian~(\ref{LOham}) is also of order $\xi$, but its size is
really about $\xi(n-n_0)^2$, which quickly can become large.  Below we
will see that $1/(2a)\approx 0.4$~MeV for tin ($Z=50$) isotopes,
$0.25$~MeV for lead isotopes $(Z=82)$, and $1.0$~MeV for $N=82$
isotones. The comparison of tin and lead isotopes on the one hand and
the $N=82$ isotones on the other hand also shows that neutron pairing
is associated with a lower energy scale than proton pairing.

In effective field theories, corrections to the leading-order are due
to neglected physics at high energy. This introduces the breakdown
energy scale $\Lambda_{\rm b}$, and a corresponding breakdown pair
number, $n_{\rm b}=\sqrt{2a\Lambda_{\rm b}}$ via Eq.~(\ref{spec}). Let
us thus assume that shell closures determine the breakdown of
pairing. Then $|n-n_0|$ cannot be larger than the number of pairs in a
major shell, i.e. $n_{\rm b}\approx 15$ or 20 in heavy nuclei.  This
would suggest that $\Lambda_{\rm b}/\xi \approx n_{\rm b}^2 \gg 1$,
and the scale separation should be large.

The subleading correction to the Lagrangian~(\ref{LOlag}) contains the
term $\dot{\alpha}^3$, and at next-to-leading order the Hamiltonian
can be written as
\be
\label{hamNLO}
H_{\rm NLO} = H_{\rm LO} + {g\over 3}\left(\hat{n}-n_0\right)^3 \ .
\ee  
Here, the factor $1/3$ is introduced for convenience. Energies are
obtained by replacing the number operator with its eigenvalues, i.e.
\be
\label{etilde}
\tilde{\varepsilon}_n = \frac{1}{2a}(n-n_0)^2 + {g\over 3}(n-n_0)^3 \ .
\ee
An estimate for the low-energy constant $g$ results from the
assumption that -- at the breakdown scale -- the correction
proportional to $g$ is clearly visible, i.e. it is as large as the
leading-order energy spacing $|E_{n_{\rm b}+1}-E_{n_{\rm b}}|\approx
|n_{\rm b}-n_0|/a$. This yields the estimate $|g|\approx
3/\left[a(n_{\rm b}-n_0)^2\right]$. To make this estimate independent
of $n_0$ we replace $(n_{\rm b}-n_0)^2$ by its average $n_{\rm
  b}^2/3$, taken over the shell. This then yields
\be
\label{b-est}
|g|\approx {9\over a n_{\rm b}^2} \ ,
\ee
and the uncertainty estimate for leading-order results is 
\be
\label{error}
\Delta \varepsilon_{n} \approx {3|n-n_0|^3 \over a n_{\rm b}^2} \ .
\ee
Below the breakdown energy, the term proportional to $g$ is then
suppressed by a factor $1/n_{\rm b} \ll 1$ compared to the leading
term.

It is clear how to generalize this approach to even higher orders: The
Lagrangian is expanded in powers of the phase velocity $\dot{\alpha}$,
and the Hamiltonian becomes an expansion in powers of $(\hat{n}-n_0)$;
subsequent orders are suppressed by increasing factors of
$\sqrt{\xi/\Lambda_{\rm b}}\sim n_{\rm b}^{-1}$.

The assumptions underlying the power counting can be tested by
extracting the low-energy coefficients $(2a)^{-1}$ and $g$ from data.
In analogy to rotations of deformed nuclei, one can also think about
subleading corrections in the framework of a variable moment of
inertia~\cite{krappe1975}. This introduces the $n$-dependent pairing
rotational constant as
\be
\label{variable}
{1\over 2} \frac{\partial^2 \tilde{\varepsilon}_n}{\partial n^2} = {1\over 2a} +g(n-n_0) \ .
\ee
This expression will be used below to extract $g$ from data.

\subsection{Odd semi-magic nuclei}
\label{sec:odd}
Pairing rotations in odd systems were previously considered in
Ref.~\cite{kishimoto1985} using a BCS state within a pairing model.
Within the effective theory they can be described as a spin-1/2
fermion coupled to the superfluid.

The Lagrangian is  
\be
\label{LOodd}
L = {a\over 2} \dot{\alpha}^2 +n_0\dot{\alpha} +L_\chi + L_{\rm int} \ .
\ee
The fermion Lagrangian is 
\be
\label{Lchi}
L_\chi =  \int {\rm d}^3\mathbf{r} \hat{\chi}^\dagger(\mathbf{r}) \left(i\partial_t +{\hbar^2\Delta_\mathbf{r}\over 2m} -V\right)\hat{\chi}(\mathbf{r}) \ , 
\ee
and the interaction $L_{\rm int}$ will be specified shortly. Here, we
have introduced the two-component fermion field
\ba
\hat{\chi}(\mathbf{r}) =\left(\begin{array}{c}
\hat{\chi}_{+{1\over 2}}(\mathbf{r})\\
\hat{\chi}_{-{1\over2}}(\mathbf{r})
\end{array}\right) \ , 
\ea
and its adjoint. The operators $\hat{\chi}_s^\dagger(\mathbf{r})$ and
$\hat{\chi}_s(\mathbf{r})$ create and annihilate a fermion with spin
projection $s=\pm 1/2$ at the position $\mathbf{r}$,
respectively. They fulfill the usual anti-commutation relations.  In
Eq.~(\ref{Lchi}) the potential is denoted as $V$ and the fermion's
mass as $m$. We neglected fermion-fermion interactions because we are
only interested in a single fermion coupled to a superfluid.

The fermion-pair number operator is 
\be
\label{ferm_num}
\hat{n}_\chi = {1\over 2}\int {\rm d}^3\mathbf{r}  \hat{\chi}^\dagger(\mathbf{r}) \hat{\chi}(\mathbf{r}) \ , 
\ee
and this operator couples the fermion to the superfluid, i.e. we have
\be
\label{interact}
L_{\rm int} = -\hat{n}_\chi \dot{\alpha} \ . 
\ee
The superfluid-fermion interaction~(\ref{interact}) is so simple
because (i) the coupling of the fermion to the superfluid must be via
the phase velocity (as we deal with a Nambu-Goldstone mode) and (ii)
it can only happen in gauge space, i.e. via the fermion-pair number
operator~(\ref{ferm_num}).  The sign is chosen for convenience.  The
canonical momentum of the superfluid is $p_\alpha = {\partial
  L\over\partial\dot{\alpha}}$, and the behavior of the superfluid
under phase transformations is as before.

Under phase transformations with an infinitesimal angle $\delta\beta$,
the fermion field changes by
\ba
e^{i\delta\beta\hat{n}_\chi} \hat{\chi}_s(\mathbf{r}) e^{-i\delta\beta\hat{n}_\chi} &=&  \hat{\chi}_s(\mathbf{r}) -i\delta\beta \hat{\chi}_s(\mathbf{r}) 
\ea
Introducing the fermion canonical momenta 
\be
\hat{\Pi}_s(\mathbf{r}) \equiv {\delta L\over\delta \partial_t\chi_s(\mathbf{r})} = i\hat{\chi}_s^\dagger(\mathbf{r}) \ ,
 \ee
and applying Noether's theorem to the coupled system then shows that
the total number of pairs
\be
\hat{n}_{\rm tot} = p_\alpha + \hat{n}_\chi 
\ee
is conserved under phase rotations. This is as expected. After the
quantization~(\ref{quant}) the eigenstates of $\hat{n}_{\rm tot}$ are
products of a superfluid state $|n\rangle$ with $n$ pairs and a
fermion state.  We denote the latter as $|q j j_z\rangle$ where $j$
denotes the fermion's total angular momentum, $j_z$ its projection
onto an arbitrary axis, and $q$ accounts for any other quantum
numbers. Thus
\be
\label{eigenodd}
\hat{n}_{\rm tot} |n\rangle |q j  j_z\rangle = \left(n + {1\over 2}\right)  |{n}\rangle |q j  j_z\rangle \ ,
\ee
and we have half integer numbers of pairs $n_{\rm tot}= n+1/2$.

A Legendre transform yields the Hamiltonian
\ba
\label{oddham}
\hat{H} &=& {1\over 2a} \left(-i\partial_\alpha + \hat{n}_\chi - n_0 \right)^2 +\hat{H}_\chi \nonumber\\
&=& {1\over 2a} \left(\hat{n}_{\rm tot} - n_0 \right)^2 +\hat{H}_\chi  \ , 
\ea
with 
\be
\hat{H}_\chi =  \int {\rm d}^3\mathbf{r} \hat{\chi}^\dagger(\mathbf{r}) \left(-{\hbar^2\Delta_\mathbf{r}\over 2m} +V\right)\hat{\chi}(\mathbf{r}) \ .
\ee
The eigenstates~(\ref{eigenodd}) of the total pair-number operator are
also eigenstates of the Hamiltonian~(\ref{oddham}).  Using
\be
\hat{H}_\chi |q j  j_z\rangle = e_{q,j} |q j j_z\rangle \ , 
\ee
we find the spectrum
\be
\label{specodd}
\varepsilon_{n_{\rm tot} q j} = {1\over 2a}\left(n_{\rm tot}-n_0 \right)^2 + e_{q j} \ .
\ee
This shows that we also have pairing rotational bands in odd-mass
nuclei. These connect states that differ by the number of pairs but
have equal spin and parity. In contrast to pairing-rotational bands in
even-even nuclei, these are not necessarily ground states. The
Hamiltonian~(\ref{oddham}) must reduce to Eq.~(\ref{hamLO}) when
acting onto the fermion vacuum. Thus, $n_0$ is an integer. Except for
the uninteresting constant $e_{q j} $ the pairing rotational bands in
odd and even nuclei have the same parabolic form. As in the case of
even isotopes, the theory for odd nuclei also predicts that pair
transfer and removal is equal in strength for states of a pairing
rotational band.  Subleading corrections are similar as for even
nuclei, i.e. we have an expansion of the Hamiltonian in powers of
$(n_{\rm tot}-n_0)$.

\subsection{Adjustment of low-energy constants}

The spectra~(\ref{spec}) and (\ref{specodd}) relate superfluid systems
in the vicinity of integer $n_0$ pairs to each other. Before we can
apply the effective theory of pairing rotations to nuclei, however, we
need to include the dominant energy contributions to nuclear
states. These consist of an overall constant and a term linear in the
number of pairs.

Let us discuss even nuclei first. Adding the contributions $E_{n_0} +
S(\hat{n}-n_0)$ to the Hamiltonian~(\ref{hamLO}) yields the energy
spectrum
\be
\label{master}
E_n = E_{n_0} - S_{n_0} (n-n_0) + \varepsilon_n \ .
\ee
Here, $E_{n_0}$ is the ground-state energy of the nucleus with $n_0$
pairs, and $S_{n_0}$ denotes the pair removal energy, and
$\varepsilon_n$ is from Eq.~(\ref{spec}). As $E_{n_0}\approx -8A$~MeV
for a nucleus with mass number $A$ and $S_{n_0}\approx 16$~MeV for
heavy nuclei, we see that the pairing rotation energies
$\varepsilon_n$ yield a small correction [except when $(n-n_0)^2\gg
  1$] because the low-energy scale $\xi$ is much smaller than
$S_{n_0}$.

The expansion~(\ref{master}) presents us with an
ambiguity~\cite{krappe1975}. One could generally argue that the
ground-state energy $E_n$ can be expanded around $n_0$ in powers of
$(n-n_0)$.  Then, our leading-order theory for pairing would be just
one contribution to the quadratic term, and other contributions are
hard to pin down without a microscopic theory. However, the
leading-order effective theory predicts that the pairing rotational
constant $a$ in Eq.~(\ref{spec}) does not depend on which nucleus
(identified by the number of pairs $n_0$) the band is centered. Within
the effective theory, any variation of $a$ must be attributed to
subleading corrections, see Eq.~(\ref{variable}). Thus, when exploring
pairing rotational bands, we can vary $n_0$ and find out if any
observed variation of $a$ is consistent with the size of subleading
contributions.

This leaves us with the following approach. We will assume that
pairing yields the dominant quadratic term in the energy expansion and
adjust the low-energy constants $E_{n_0}$, $S_{n_0}$, and $a$ to the
binding energies of the nuclei with $n_0$ and $n_0\pm 1$ pairs.  We
use
\ba
\label{sol1}
S_{n_0} &=& \frac{1}{2}\left(E_{n_0-1}-E_{n_0+1}\right) \ , \nonumber\\
a^{-1} &=& E_{n_0+1}-2E_{n_0}+E_{n_0-1}\ .
\ea
We see that $S_{n_0}$ is the average of two two-nucleon separation
energies, while the rotational constant is a three-point difference of
even nuclei.  Clearly, when adjusting $a$ this way it becomes an
$n_0$-dependent quantity, and the variations of $a$ with $n_0$ inform
us about the size of subleading corrections.

\begin{figure}[!htbp]
    \includegraphics[width=0.99\linewidth]{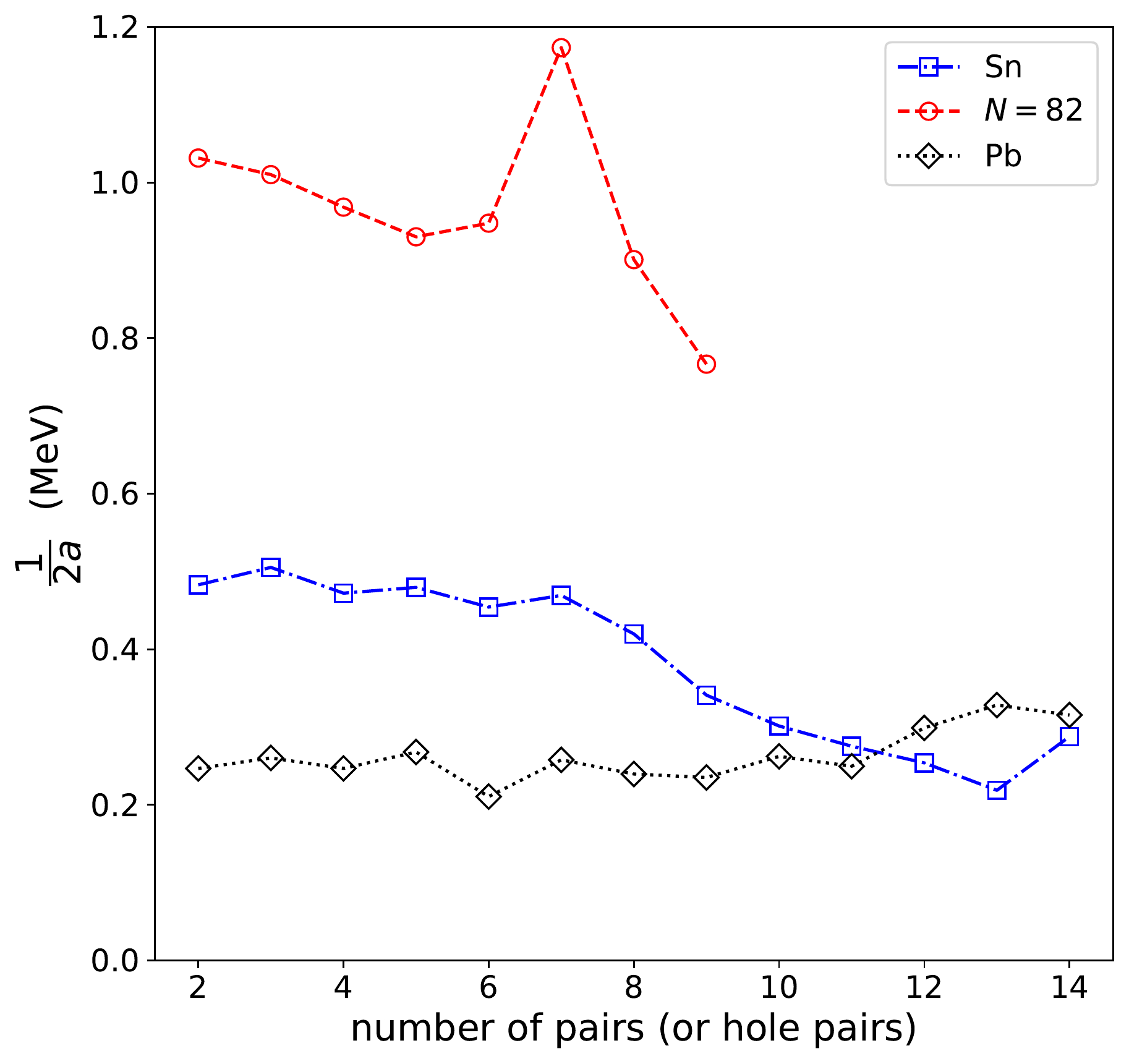}
    \caption{Pairing rotational constants for even isotopes of tin (as
      a function of pairs above neutron number $N=50$), of $N=82$
      isotones (as function of pairs above proton number $N=50$) and
      lead (as a function of pair holes below the neutron number
      $N=126$).}
    \label{fig:arot}
\end{figure}

Figure~\ref{fig:arot} shows the pairing rotational constant
$(2a)^{-1}$, computed via Eq.~(\ref{sol1}), for even isotopes of tin
(as a function of pairs above neutron number $N=50$), of $N=82$
isotones (as function of pairs above proton number $N=50$) and lead
(as a function of pair holes below the neutron number $N=126$). We see
that the pairing rotational constant is approximately $n_0$
independent for isotopes of lead while the $N=82$ isotones and the
isotopes of tin exhibit more variations. This suggests that
higher-order corrections are significant in those nuclei. We also see
that the variations are not smooth as the number of pairs (or pair
holes in lead) changes. This suggests that the (smooth) subleading
contributions discussed in Sect.~\ref{sec:power} are only part of the
corrections beyond quadratic order. The non-smooth fluctuations are
outside the scope of the effective theory. They also prevent us from
adjusting subleading low-energy constants locally, i.e. in a vicinity
of a given $n_0$.

Therefore, let us consider global adjustments of $g$ in
Eq.~(\ref{etilde}) and check the power counting. The average slopes of
the lines in Fig.~\ref{fig:arot} are small compared to the rotational
constants, and this suggests that the smooth subleading correction
could be systematic. We can use Eq.~(\ref{variable}) and identify the
average slope as $g$. Table~\ref{tab1} presents the average values of
the rotational constant $ (2a)^{-1}$ and $g$ for the tin and lead
isotopes and the $N=82$ isotones. Also shown is the maximum number of
pairs for the major shell corresponding to the nuclei of interest, and
the estimate $3/(\langle a\rangle n_{\rm b}^2)$ from Eq.~(\ref{b-est})
for the size of the coupling $g$. The theoretical estimates correctly
identify the scale of $\langle|g|\rangle$ (they are about twice of
what was extracted from data), and this gives us confidence in the
power counting proposed in Sect.~\ref{sec:power}.

\begin{table}
\caption{\label{tab1}Average values of pairing rotational constants
  $(2a)^{-1}$ and the absolute average scale $|\langle g\rangle |$ for
  subleading correction (both in MeV) for isotopes of tin and lead,
  and $N=82$ isotones. Also shown are the maximum number of pairs
  $n_b$ in the relevant major shell, and -- in the last column -- the
  estimate~(\ref{b-est}) for the size of the low-energy constant $g$
  (in MeV).}
\begin{ruledtabular}
\begin{tabular}{| l | c | l | l | l | }
   &  $\langle (2a)^{-1}\rangle$  & $|\langle g\rangle|$ & $n_{\rm b}$ &  $9/(\langle a\rangle n_{\rm b}^2)$ \\\hline
  Sn         & 0.38 & 0.016   & 16 & 0.027 \\
  Pb         & 0.26 & 0.0057 & 22 & 0.0097\\
  $N=82$ & 0.97 & 0.038   & 16 & 0.068\\ 
\end{tabular}
\end{ruledtabular}
\end{table}

Thus, the uncertainty estimate~(\ref{error}) is expected to capture
the smooth corrections to the leading-order pairing rotational
bands. In what follows we will assume that pairing does yield the
dominant quadratic contribution to the expansion~(\ref{master}), limit
the discussion to the leading-order theory, and use the uncertainty
estimate~(\ref{error}).

For odd nuclei, we expand the pairing rotational contribution as
$(n_{\rm tot} -n_0)^2 = (n_{\rm tot}-n_0-1/2)^2 +[n_{\rm
    tot}-n_0-1/4]$. The constant and linear terms $[n_{\rm
    tot}-n_0-1/4]$ can then be absorbed in an expansion of the
energy~(\ref{master}). Thus, we will employ Eqs.~(\ref{master}) and
(\ref{sol1}) for even nuclei (by using integer $n_0$) and for odd
nuclei (by using half integer $n_0$). Inspection shows that the
pairing rotational constants for the odd nuclei are close to their
even neighbors. This allows us to use the data in Table~\ref{tab1}
also for uncertainty estimates in odd nuclei.


\section{Effective theory for two superfluids}
\label{sec:two}

\subsection{Even-even nuclei}
In heavy open-shell nuclei, neutrons form isovector pairs and so do
protons, and both superfluids interact. Thus, we do not consider
proton-neutron pairing but will include interactions between proton
and neutron pairs. The effective theory is based on the emergent
symmetry breaking from $U(1)\times U(1)\to 1$, and the coset is
isomorph to $U(1)\times U(1)$. The phases $\alpha(t)$ and $\beta(t)$
denote the Nambu-Goldstone modes corresponding to neutron and proton
pairs, respectively. The most general Lagrangian up to quadratic terms
in phase velocities is
\ba
L = {1\over 2}(\dot{\alpha}, \dot{\beta}) 
\hat{M}
\left(
\begin{array}{c}
\dot{\alpha}\\
\dot{\beta}
\end{array}
\right)
 + (n_0, z_0)
 \left(
\begin{array}{c}
\dot{\alpha}\\
\dot{\beta}
\end{array}
\right) \ .
\ea
Here, $n_0$ and $z_0$ are low-energy constants and $\hat{M}$ is a
symmetric $2\times 2$ ``mass'' matrix with three parameters, and we
employed a matrix-vector notation. The off-diagonal elements of
$\hat{M}$ introduce an interaction between the two
superfluids. Introducing the canonical momenta $p_\alpha\equiv
{\partial L\over\partial\dot{\alpha}}$ and $p_\beta\equiv {\partial
  L\over\partial\dot{\beta}}$, and performing a Legendre transform
yields the Hamiltonian
\ba
H = {1\over 2}(p_{\alpha}-n_0, p_{\beta}-z_0) 
\hat{M}^{-1}
\left(
\begin{array}{c}
p_{\alpha}-n_0\\
p_{\beta}-z_0
\end{array}
\right) \ .
\ea
Quantization proceeds as in the previous Section, and single
valuedness of the wavefunction under simple gauge transformations
requires that $n_0$ and $z_0$ are integers.  The resulting Hamiltonian
is
\ba
\label{ham2}
H = {1\over 2}(\hat{n}-n_0, \hat{z}-z_0) 
\hat{M}^{-1}
\left(
\begin{array}{c}
\hat{n}-n_0\\
\hat{z}-z_0
\end{array}
\right) \ , 
\ea
where $\hat{n}\equiv -i\partial_\alpha$ and $\hat{z}\equiv
-i\partial_\beta$ count the conserved number of pairs in each
superfluid. Energies are obtained by replacing these number operators
by their eigenvalues, i.e.
\be
\label{spec2}
\varepsilon_{n, z} = {1\over 2a}(n-n_0)^2 +{1\over 2b}(z-z_0)^2 +{1\over c}(n-n_0)(z-z_0) \ .
\ee
Here, we have chosen the constants $1/a$, $1/b$, and $1/c$ as the
diagonal and off-diagonal entries of $\hat{M}^{-1}$, respectively. We
see that all even-even nuclei in an entire region are connected via
pairing, and the spectrum is an elliptical paraboloid; any section of
this paraboloid is a pairing rotational band, and the sections are not
limited to keeping neutron or proton numbers fixed.

It is interesting to use the isospin projection $T=n+z$ and mass
number $A=2(n+z)$ as independent variables (and similarly introduce
$T_0$ and $A_0$). Then the spectrum~(\ref{spec2}) becomes
\ba
\label{specTA}
\varepsilon(T,A) &=& \left({1\over 2a}+{1\over 2b}-{1\over c}\right) \left({T-T_0\over 2}\right)^2 \nonumber\\
&+& \left({1\over 2a}+{1\over 2b}+{1\over c}\right) \left({A-A_0\over 4}\right)^2\nonumber\\
&+&\left({1\over a}-{1\over b}\right) {(T-T_0)(A-A_0)\over 8} \ .
 \ea

The spectrum~(\ref{spec2}) recovers the results of
Refs.~\cite{beck1972,krappe1975,marshalek1977,hinohara2015}.  The
effective theory thus supports the recent proposal by
\textcite{hinohara2016} to employ the pairing rotational tensor
$\hat{M}^{-1}$ as a model-independent indicator for pairing. Its
eigenvectors are expected to point into the directions of the valley
of $\beta$ stability and perpendicular to it; the corresponding
eigenvalues are expected to be small and large in magnitude,
respectively.

The eigenstates of the Hamiltonian~(\ref{ham2}) are product states
$|n, z\rangle$ that specify the number of pairs in each fluid, i. e.
\ba
\hat{n}|n,z\rangle &=&  n |{n},z\rangle \ , \nonumber\\
\hat{z}|n,z\rangle &= & z |n,z\rangle \ .
\ea
Analogous to the case of one superfluid [see Eq.~(\ref{pair})] we can
introduce pair removal (or pair addition) operators for each
superfluid via
\ba
\hat{P} &=&  P_0 e^{-i\alpha} \ , \nonumber\\
\hat{Q} &=&  Q_0 e^{-i\beta} \ .
\ea
We then have
\ba
\hat{P}|n,z\rangle &=&  P_0 |{n-1},z\rangle \ , \nonumber\\
\hat{Q}|n,z\rangle &=&  Q_0 |n,z-1\rangle \ .
\ea
Double charge exchange reactions are then governed by the nuclear
matrix elements
\ba
\langle n-1,z+1|\hat{P}\hat{Q}^\dagger|n,z\rangle &=&  P_0 Q_0^* \ , \nonumber\\
\langle n+1,z-1|\hat{P}^\dagger\hat{Q}|n,z\rangle &=&  P_0^* Q_0  \ ,
\ea
while the transfer or removal of $\alpha$ particles involves the
nuclear matrix elements
\ba
\langle n-1,z-1|\hat{P}\hat{Q}|n,z\rangle &=&  P_0 Q_0 \ , \nonumber\\
\langle n+1,z+1|\hat{P}^\dagger\hat{Q}^\dagger|n,z\rangle &=&  P_0^* Q_0^*  \ .
\ea
We thus see that the leading-order theory of pairing predicts that
four different reactions involve the same absolute squared nuclear
matrix element, which is independent of $n$ and $z$. As pair transfer
in single superfluid systems, these are testable predictions for two
coupled superfluids.

Let us briefly discuss subleading corrections of the
Hamiltonian~(\ref{ham2}). These are in powers of
$(\hat{n}-n_0)^k(\hat{z}-z_0)^l$ with $k+l=3$. Alternatively, and with
view on Eq.~(]\ref{specTA}), one could also include powers
  $(T-T_0)^k(A-A_0)^l$. Following the steps in Sect.~\ref{sec:power}
  that led to Eq.~(\ref{b-est}) we can also here estimate the
  uncertainties and find
\be
\begin{aligned}
\label{error2}
\Delta \varepsilon_{n,z_0} &\approx {3|n-n_0|^3\over a n_{\rm b}^2} \ , \\ 
\Delta \varepsilon_{n_0,z} &\approx {3|z-z_0|^3\over b  z_{\rm b}^2} \ , \\
\Delta \varepsilon(T,A_0) &\approx {3\over 4}\left({1\over 2a}+{1\over 2b}-{1\over c}\right) \frac{|T-T_0|^3}{{\rm min}(z_b^2,n_b^2)} \ ,  \\
\Delta \varepsilon(T_0,A) &\approx {3\over 32}\left({1\over 2a}+{1\over 2b}+{1\over c}\right) \frac{|A-A_0|^3}{{\rm min}(z_b^2,n_b^2)} \ ,
\end{aligned}
\ee
for pairing rotational bands in isotopes, isotones, isobars, and
nuclei with the isospin projection, respectively, of the nucleus with
$n_0$ neutron and $z_0$ proton pairs.

The effective field theory can also be extended to odd and to odd-odd
nuclei, and one can easily write down the leading-order result.
However, in practical applications, it is difficult to trace how
states with non-zero spins evolve as neutron and proton numbers are
changed, and this is particularly so for odd-odd nuclei. For this
reason, such extensions of the theory are not pursued in this paper.

\subsection{Adjustment of low-energy constants}
As was the case for a single superfluid, we have to add the dominant
contributions $E_{n_0, z_0} - S_{n_0}(\hat{n}-n_0)
-S_{z_0}(\hat{z}-z_0)$ to the Hamiltonian~(\ref{ham2}) and find the
energy spectrum
\ba
\label{master2}
E_{n,z}&=&E_{n_o,z_0} - S_{n_0}(n-n_0) -S_{z_0}(z-z_0) +\varepsilon_{n,z}  \ . \nonumber\\
\ea
Here, $\varepsilon_{n,z}$ is from Eq.~(\ref{spec2}) and contains the
low-energy constants $a$, $b$, while $c$, and $S_{n_0}$ and $S_{z_0}$
are (approximately) pair separation energies.  We adjust the
parameters $S_{n_0}$ and $a$ (and $S_{z_0}$ and $b$) similarly as in
the case of a single superfluid [see Eq.~(\ref{sol1})] and have
\ba
\label{sol2}
S_{n_0} &=& \frac{1}{2}\left(E_{n_0-1,z_0}-E_{n_0+1,z_0}\right) \ , \nonumber\\
a^{-1} &=& E_{n_0+1,z_0}-2E_{n_0,z_0}+E_{n_0-1,z_0} \ , \nonumber\\
S_{z_0} &=& \frac{1}{2}\left(E_{n_0,z_0-1}-E_{n_0,z_0+1}\right) \ , \nonumber\\
b^{-1} &=& E_{n_0,z_0+1}-2E_{n_0,z_0}+E_{n_0,z_0-1}\ .
\ea
We need one more datum to determine $c$ and choose the symmetric
expression
\ba
\label{solc}
c^{-1} =\frac{1}{4}&&\big(E_{n_0+1,z_0+1}-E_{n_0+1,z_0-1}\nonumber\\
&&+E_{n_0-1,z_0-1}-E_{n_0-1,z_0+1}\big)  \ .
\ea
%


\section{Applications}
\label{sec:app}
\subsection{Single superfluid: semi-magic nuclei}

\begin{figure}[!htbp]
    \includegraphics[width=0.99\linewidth]{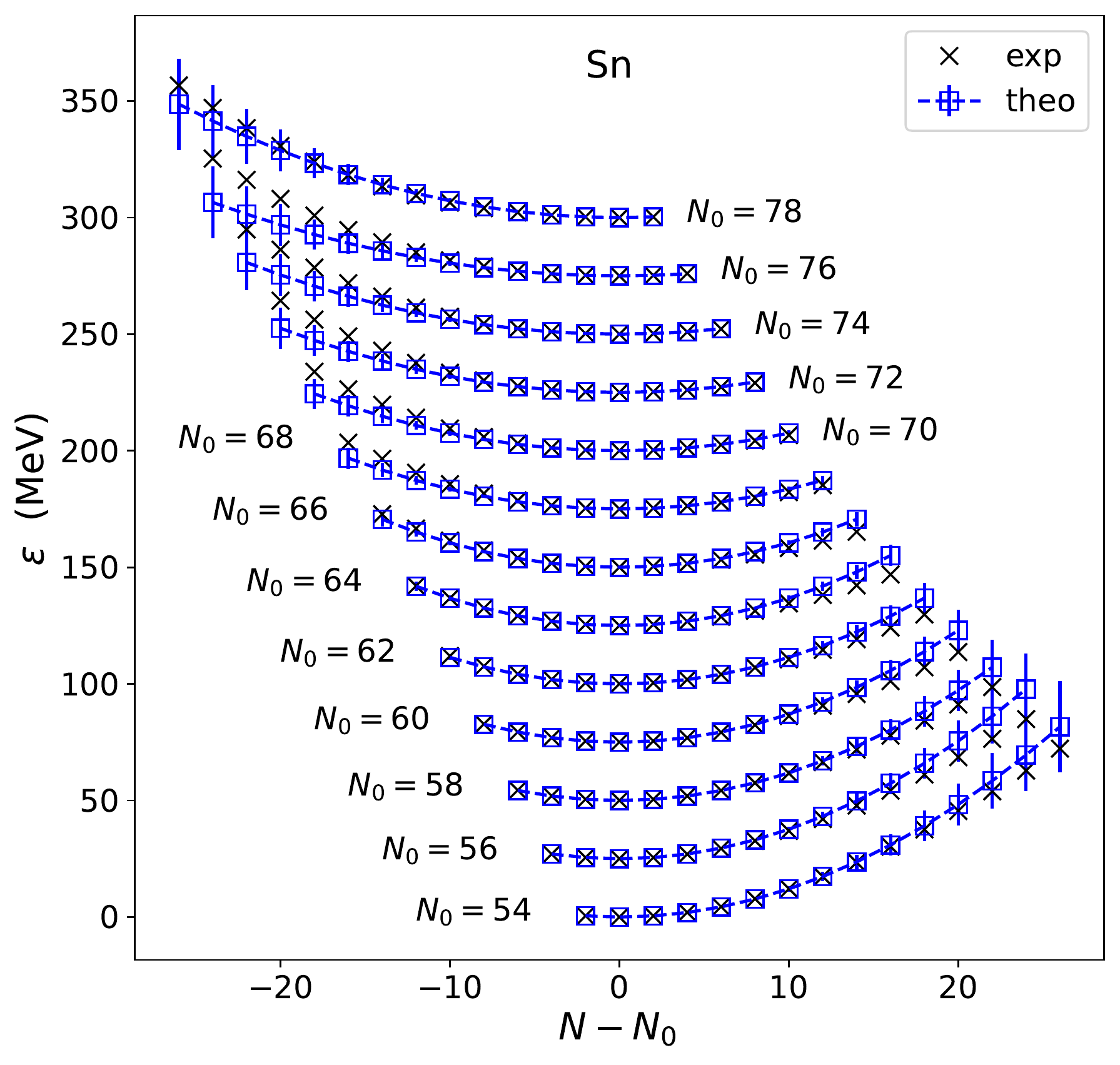}
    \caption{Pairing rotational bands in tin isotopes, centered on
      nuclei with $N_0$ neutrons as indicated: Experimental data $E_n
      - E_{n_0} + S_{n_0}(n-n_0)$ is compared with the theory
      prediction $\varepsilon_n=\left(n -n_0\right)^2/(2a)$ for nuclei
      with $n$ pairs around $n_0$. Errorbars are uncertainty estimates
      from omitted subleading terms. Bands are shifted by multiples of
      25~MeV as $N_0=2n_0$ is increased from 54 to 78.  In each band,
      the energies with $|N-N_0|\le 2$ have been adjusted to data.}
    \label{fig:Sn}
\end{figure}

Figure~\ref{fig:Sn} shows the pairing rotational band in tin isotopes
centered on neutron number $N_0$ as indicated. Different bands are
shifted by 25~MeV as $N_0$ is increased from 54 to 78. The number of
pairs is $n=N/2$ and $n_0=N_0/2$. Experimental data $E_n - E_{n_0} +
S_{n_0} (n-n_0)$ is compared with the theory prediction
$\varepsilon_n$, see Eq.~(\ref{master}).  Here and in what follows,
the $y$-axis is simply labelled as $\varepsilon$. Errorbars show the
uncertainty estimates~(\ref{error}) using the average value of $a$
from Table~\ref{tab1}.  We see that theory describes data accurately
within errorbars. For each band, the three lowest-energy points with
$-2\le N-N_0\le 2$ have been adjusted to data.

\begin{figure}[!htbp]
    \includegraphics[width=0.99\linewidth]{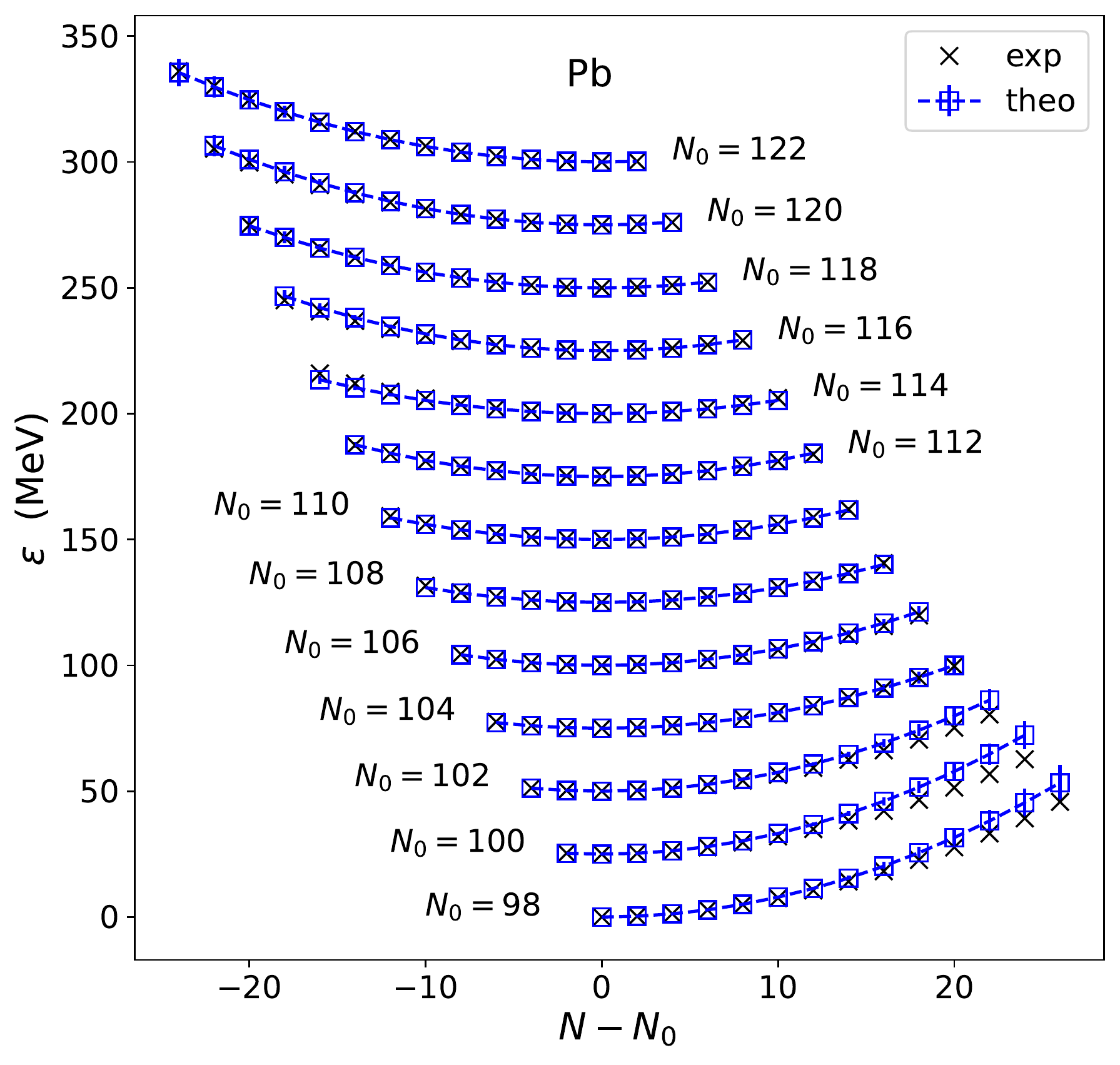}
    \caption{Pairing rotational bands in lead isotopes centered on
      nuclei with neutron number $N_0$ as indicated: Experimental data
      $E_n - E_{n_0} + S_{n_0}(n-n_0)$ is compared with the theory
      prediction ${1\over 2a}\left(n -n_0\right)^2$ for nuclei with
      $n$ pairs around $n_0$.  Errorbars are uncertainty estimates for
      omitted subleading terms. Bands are shifted by multiples of
      25~MeV as $N_0=2n_0$ is increased from 98 to 122.  In each band,
      the energies with $|N-N_0|\le 2$ have been adjusted to data.}
    \label{fig:Pb}
\end{figure}

Figure~\ref{fig:Pb} shows the pairing rotational bands in lead
isotopes centered on neutron number $N_0$ as indicated. Different
bands are shifted by 25~MeV as $N_0$ is increased from 98 to
122. Errorbars again show the uncertainty estimates~(\ref{error})
using the average value of $a$ from Table~\ref{tab1}. Theory describes
data accurately within the uncertainty estimates.

\begin{figure}[!htbp]
    \includegraphics[width=0.99\linewidth]{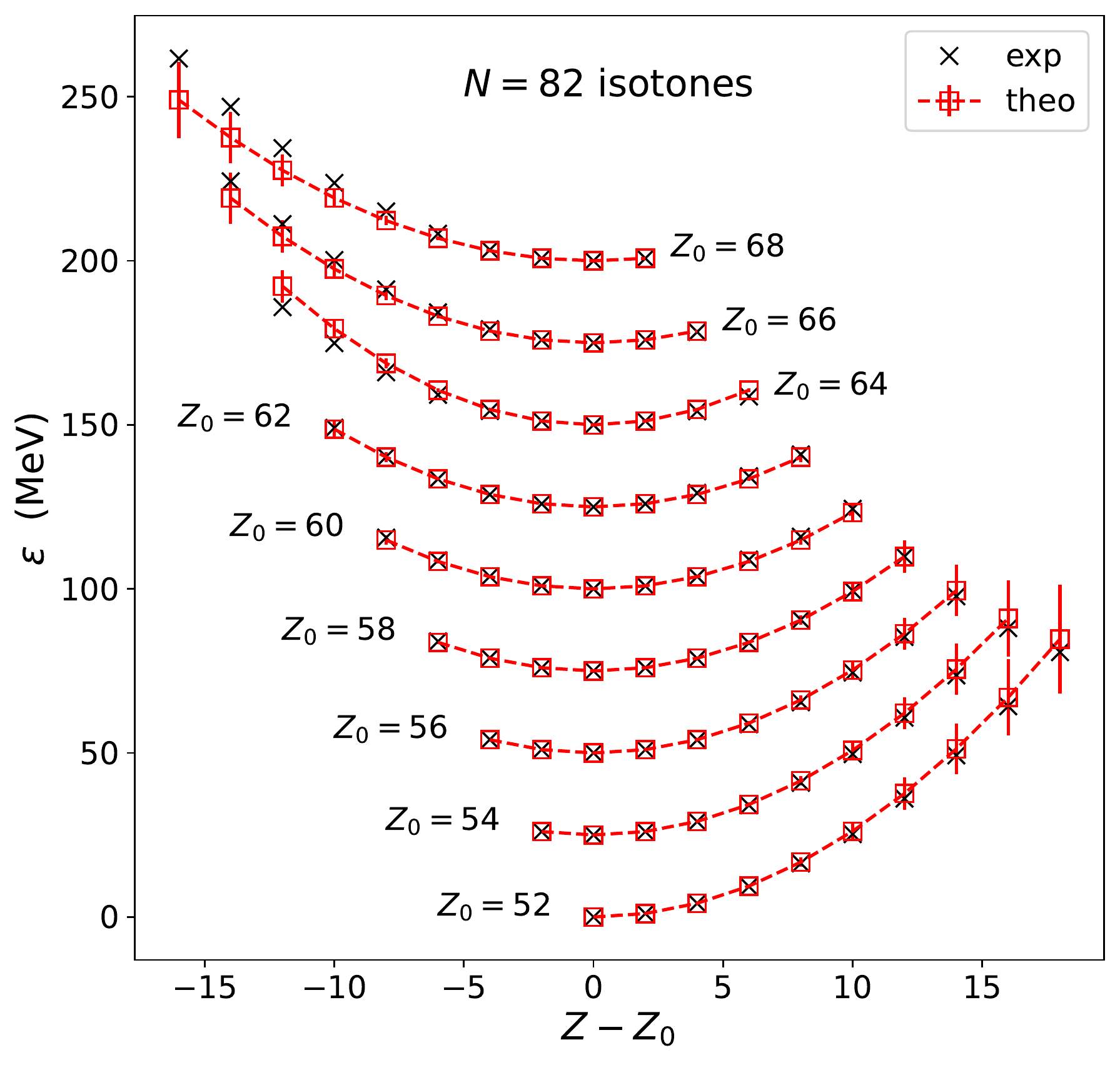}
    \caption{Pairing rotational bands in $N=82$ isotones, centered on
      nuclei with proton number $Z_0$ as indicated. Experimental data
      $E_n - E_{n_0} + S_{n_0}(n-n_0)$ is compared with the theory
      prediction ${1\over 2a}\left(n -n_0\right)^2$ for nuclei with
      $n$ pairs around $n_0$. Uncertanties estimate the omitted
      contributions from subleading terms. Bands are shifted by 25~MeV
      as $Z_0=2n_0$ is increased from 52 to 68.  In each band, the
      energies with $|Z-Z_0|\le 2$ have been adjusted to data.}
    \label{fig:N82}
\end{figure}

Figure~\ref{fig:N82} shows the pairing rotational bands in $N=82$
isotones centered on nuclei with proton number $Z_0$ as
indicated. Different bands are shifted by 25~MeV as $Z_0$ is increased
from 52 to 68. The number of pairs is $n=Z/2$ and $n_0=Z_0/2$.
Uncertainty estimates are based on Eq.~(\ref{error}) and the value of
$a$ from Table~\ref{tab1}. Again, theory describes data accurately
within uncertainties.
 
In summary, the leading-order Hamiltonian~(\ref{hamLO}) yields an
accurate description of pairing rotational bands within uncertainty
estimates. This gives confidence in the power counting and the
underlying separation of scales in even semi-magic nuclei.

\subsection{Odd semi-magic nuclei}
Let us also test the effective field theory prediction for odd
nuclei. The ground-state spin typically evolves across an isotopic or
isotonic chain, and we focus therefore on low-lying states with
constant spin and parity. The excitation energy of such states must be
added to the ground-state energies $E_n$ in Eq.~(\ref{sol1}) when
computing the low-energy constants.

\begin{figure}[!htbp]
    \includegraphics[width=0.99\linewidth]{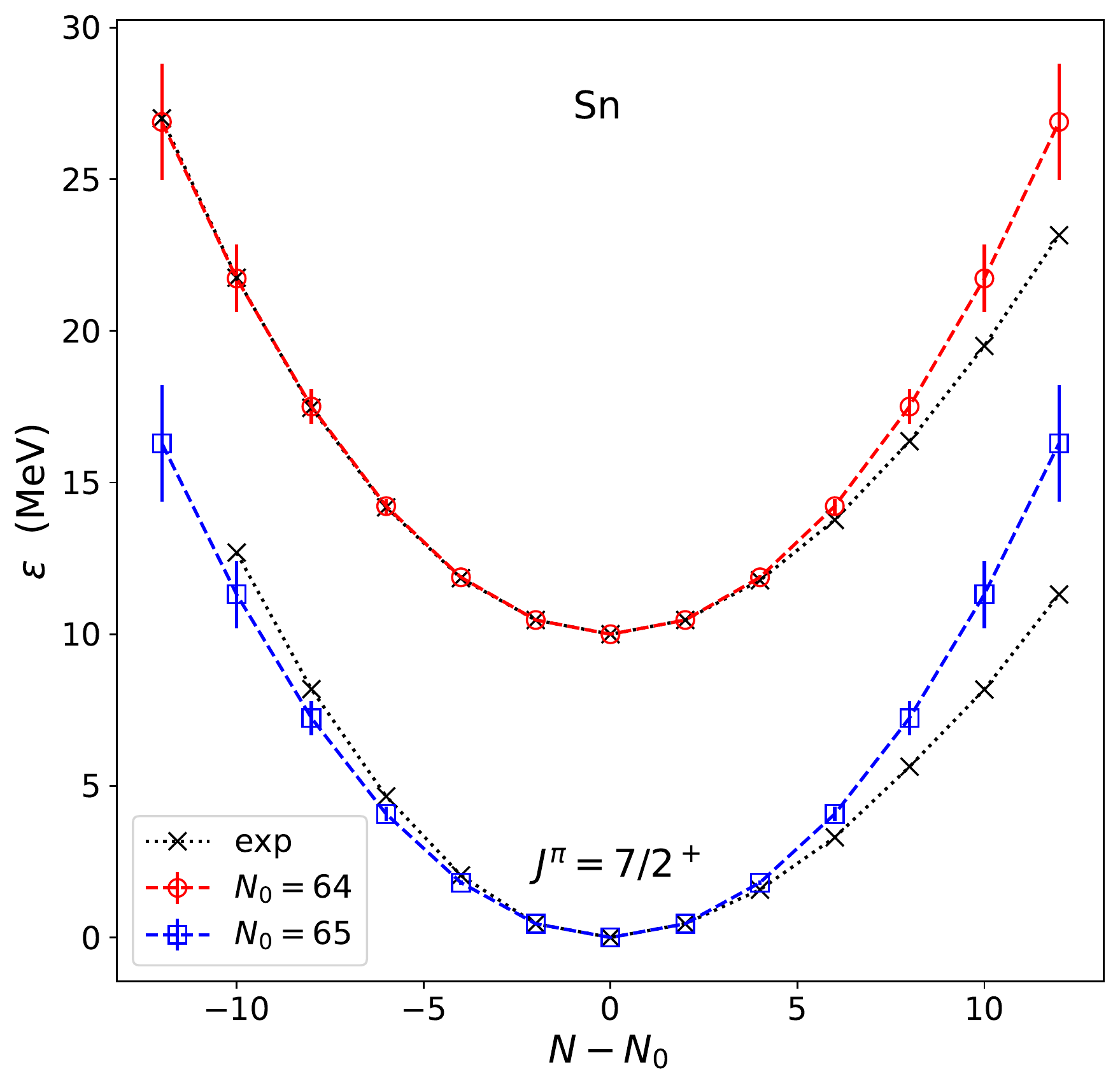}
    \caption{Pairing rotational bands in odd (blue squares) and even
      (red circles) tin isotopes.  The odd nuclei have spin/parity
      $J^\pi=7/2^+$ with $^{115}$Sn ($N_0=65$) at the center, while
      the even nuclei are centered at $^{114}$Sn. Data is shown as
      black crosses. In each band, the central three points are
      adjusted to data.}
    \label{fig:oddSn}
\end{figure}

In the odd tin isotopes we focus on the $J^\pi=7/2^+$ state that is
low in energy and compute the pairing rotational band for the nucleus
with neutron number $N_0=65$. The results are shown in
Fig.~\ref{fig:oddSn} and compared to a pairing rotational band in the
neighboring even isotopes (centered at $N=64$ and shifted by 10~MeV).
The uncertainty estimates~(\ref{error}) with $a$ from Table~\ref{tab1}
reflect the scale of deviations from data but do not capture them
quantitatively for the larger values of $N-N_0$.

\begin{figure}[!htbp]
    \includegraphics[width=0.99\linewidth]{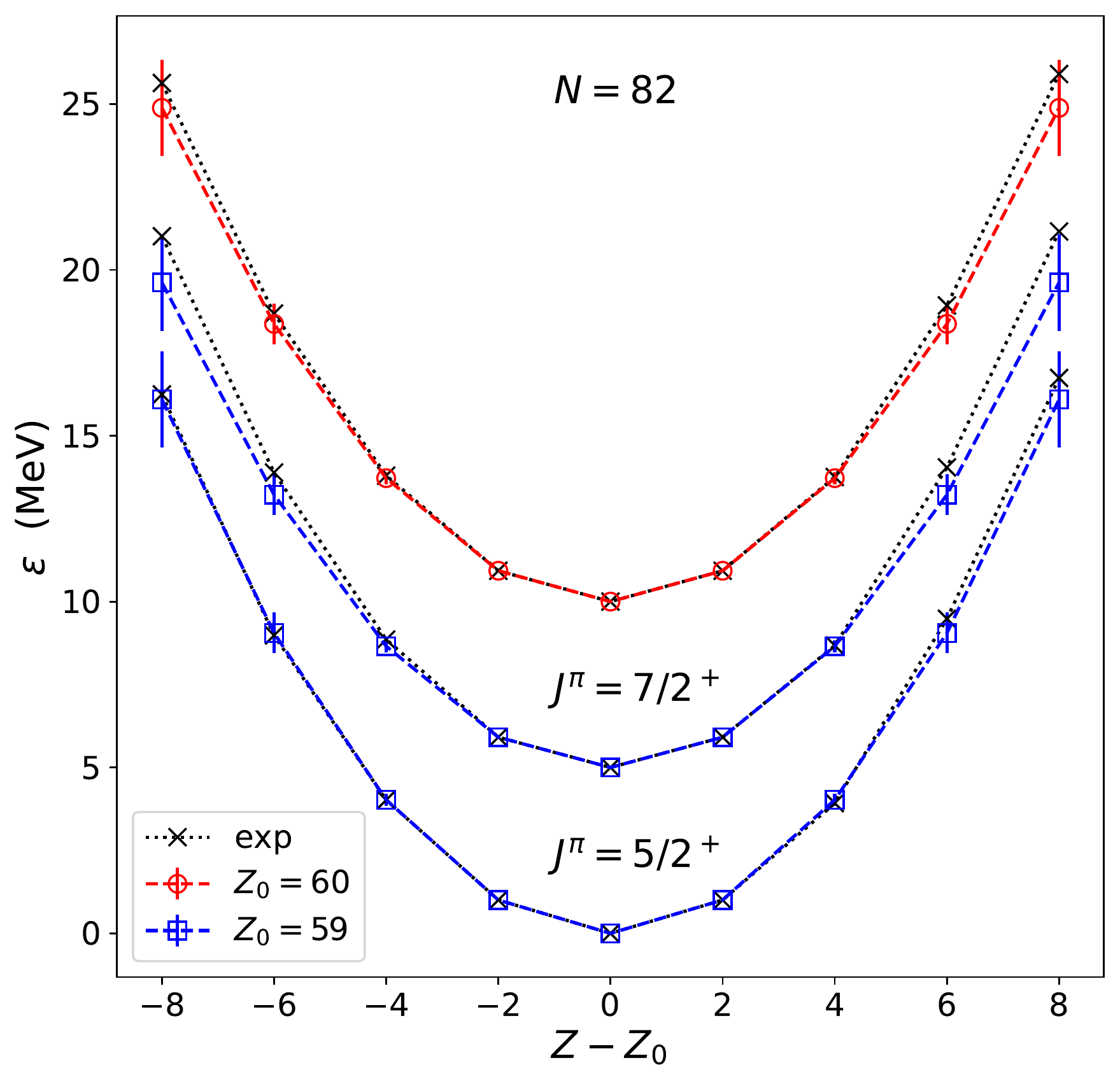}
    \caption{Pairing rotational bands in odd (blue squares) and even
      (red circles) $N=82$ isotones.  Two bands connecting odd nuclei
      in with spin/parity as indicated are centered on the Pr nucleus
      ($Z=59$) and compared to the even ground-state band with the Nd
      nucleus ($Z=60$) at its center. Data are shown as black
      crosses. In each band, the central three points are adjusted to
      data. Bands are shifted by multiples of 5~MeV.}
    \label{fig:odd82}
\end{figure}

The agreement between theory and experiment is better in $N=82$
isotones.  In the odd isotones we focus on the $J^\pi=5/2^+$ and
$7/2^+$ states that are low in energy and can easily be traced across
the chain, taking $Z=59$ (element Pr) as the central nucleus of the
pairing rotational band. The results are shown in Fig.~\ref{fig:odd82}
and compared to the pairing rotational band in even isotones, centered
at the Nd nucleus $(Z=60)$. The uncertainty estimate~(\ref{error})
uses the value of $a$ from Table~\ref{tab1} and captures the
differences between theory and data.

\begin{figure}[!htbp]
    \includegraphics[width=0.99\linewidth]{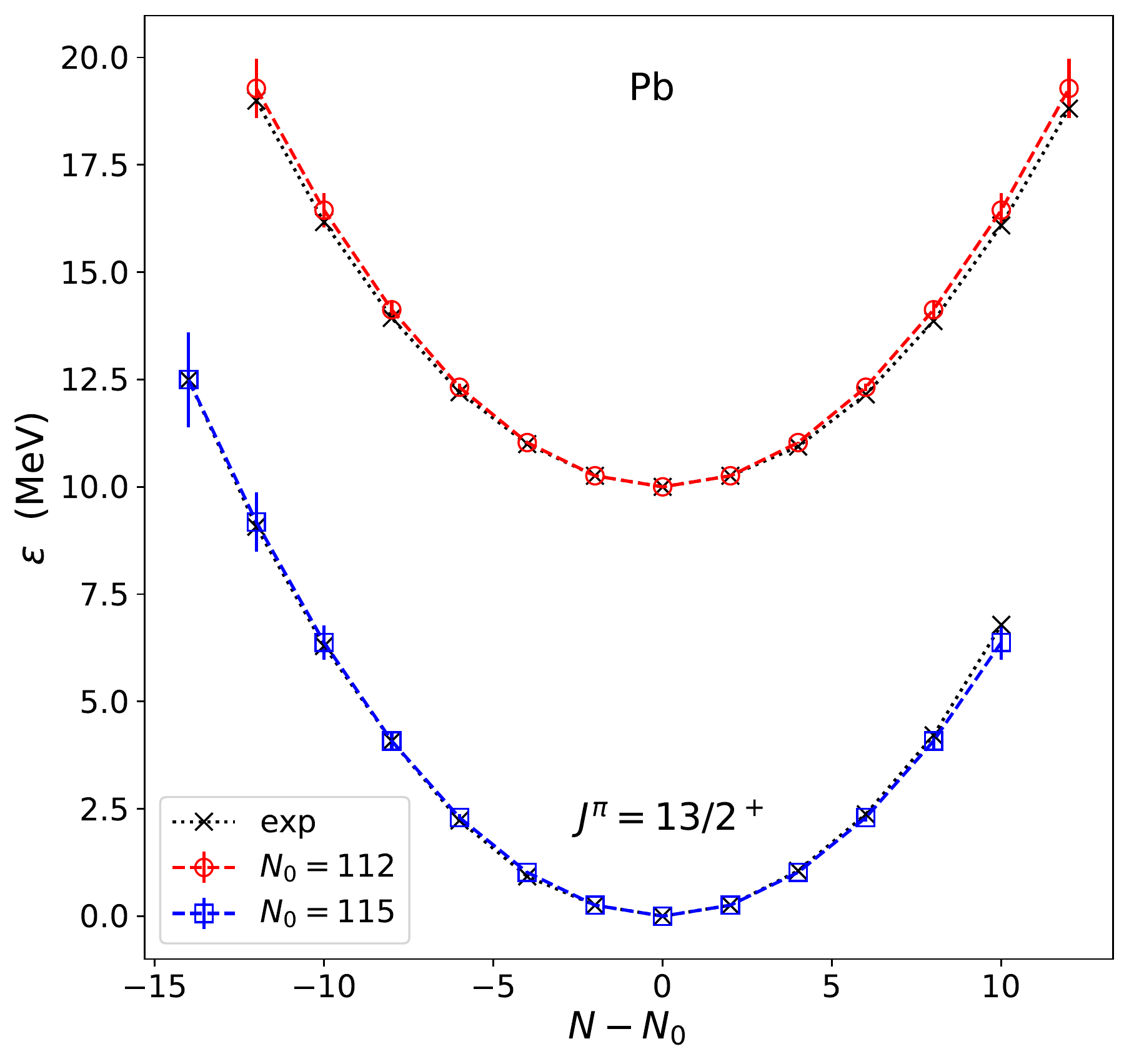}
    \caption{Pairing rotational bands in odd (blue squares) and even
      (red circles) lead isotopes.  The bands connecting odd nuclei
      with spin/parity $J^\pi=13/2^+$ are centered on $^{197}$Pb and
      compared to the even ground-state band with $^{112}$Pb at its
      center. Data are shown as black crosses. In each band, the
      lowest three points are adjusted to data. Bands are shifted by
      10~MeV.}
    \label{fig:oddPb}
\end{figure}

Finally we turn to lead. Here, an isomeric $J^\pi=13/2^+$ state is
known in odd isotopes lighter than $^{208}$Pb, although its exact
spacing with respect to the ground state is only known for $^{195}$Pb
and heavier isotopes; we use tentative spin assignments for more
neutron-deficient isotopes and take $^{197}$Pb as the center for the
computation of the pairing rotational band.  The results are shown in
Fig.~\ref{fig:oddPb} and compared to an even isotope. The uncertainty
estimate~(\ref{error}) with $a$ from Table~\ref{tab1} captures the
discrepancies between data and theory.

Overall, the results of this Subsection show that the effective field
theory also delivers accurate results for pairing rotational bands in
odd semi-magic nuclei. We also see that the pairing rotational bands
for even and odd semi-magic nuclei have the same pairing rotational
constant to a very good approximation.

\subsection{Two superfluids: open-shell nuclei}
\label{sec:app2}

Let us take $^{166}$Yb as the $(Z_0=70, N_0=96)$ nucleus in the center
of the rare-earth region and adjust the low-energy constants from
Eqs.~(\ref{sol2}) and (\ref{solc}) to its immediate even-even
neighbors. This yields the pairing rotational constants $1/(2a)\approx
0.30$~MeV, $1/(2b)\approx 0.94$~MeV, $1/c\approx -0.95$~MeV. The
proton and neutron pairing rotational constants are consistent with
those presented in Table~\ref{tab1} for Pb isotopes and $N=82$
isotones, respectively. The size of the off-diagonal coupling $1/c$
shows that the interaction of the two superfluids is
strong~\cite{wang2014,hinohara2015,hinohara2016}. The curvature is
small for pairing at constant isospin projection and large for isobars
[see Eq.~(\ref{specTA})].

\begin{figure}[!hbt]
    \includegraphics[width=0.99\linewidth]{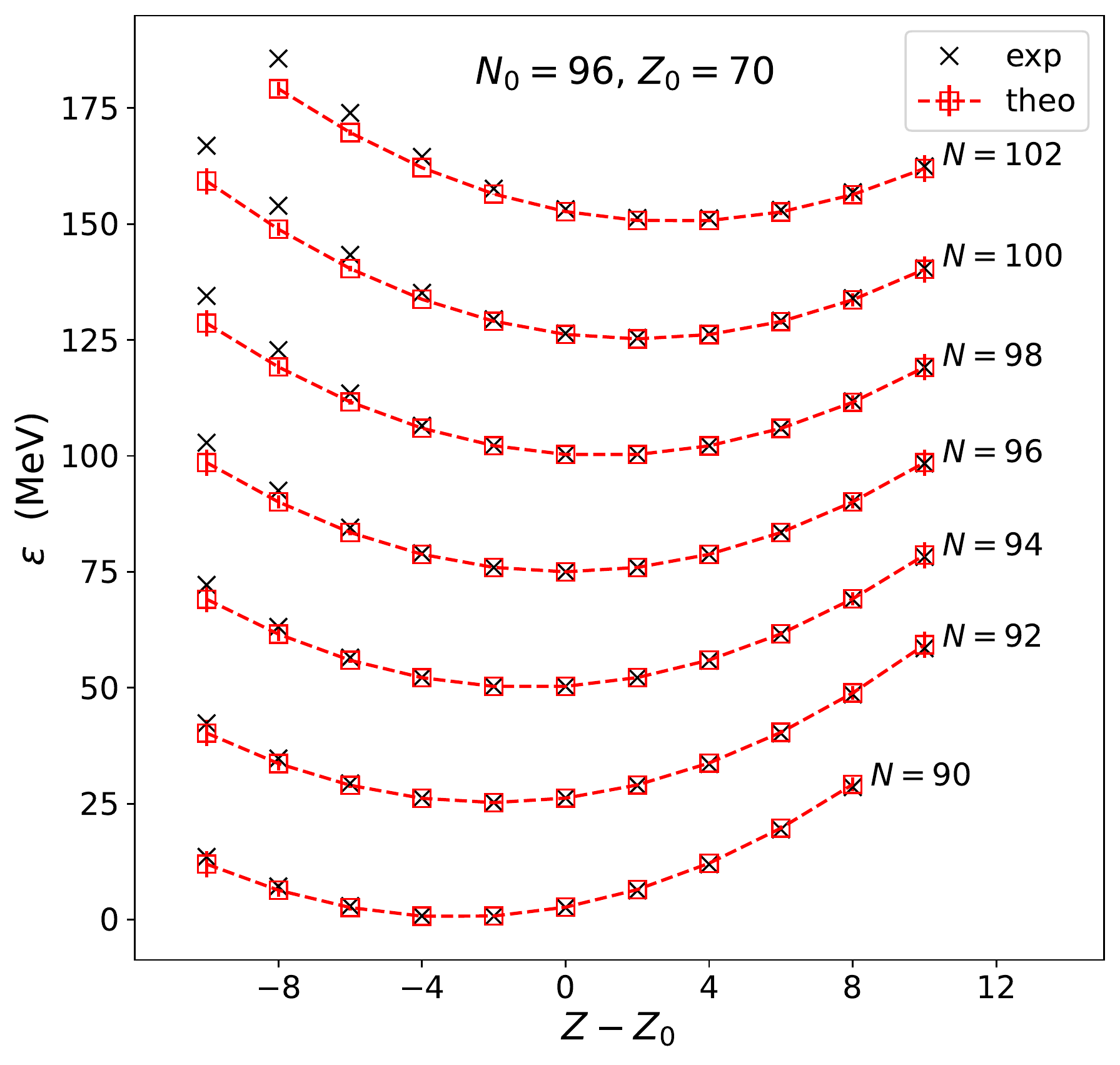}
    \caption{Proton-pairing rotational bands as sections of a pairing
      elliptical paraboloid. Bands for neutron numbers as indicated in
      the rare earth region around $^{166}$Yb ($Z_0=70, N_0=96$):
      Experimental data is compared with the theory prediction. A
      total of six low-energy constant has been adjusted for all shown
      bands. Different bands are shifted by multiples of 12~MeV.}
    \label{fig:Yb166-Zpairs}
\end{figure}

Figure~\ref{fig:Yb166-Zpairs} shows the proton-pairing rotational
bands (shifted by multiples of 12~MeV) for fixed neutron number $N$.
Overall, theory and data agree reasonably well, and only for large
values of $|Z-Z_0|$, and significant away from $N=96$ do we see
disagreement.  The error estimates are based on
$\Delta\varepsilon_{n_0,z}$ from Eqs.~(\ref{error2}). They are too
small to capture the deviations for large $N$ and small $Z$.

\begin{figure}[!htb]
    \includegraphics[width=0.99\linewidth]{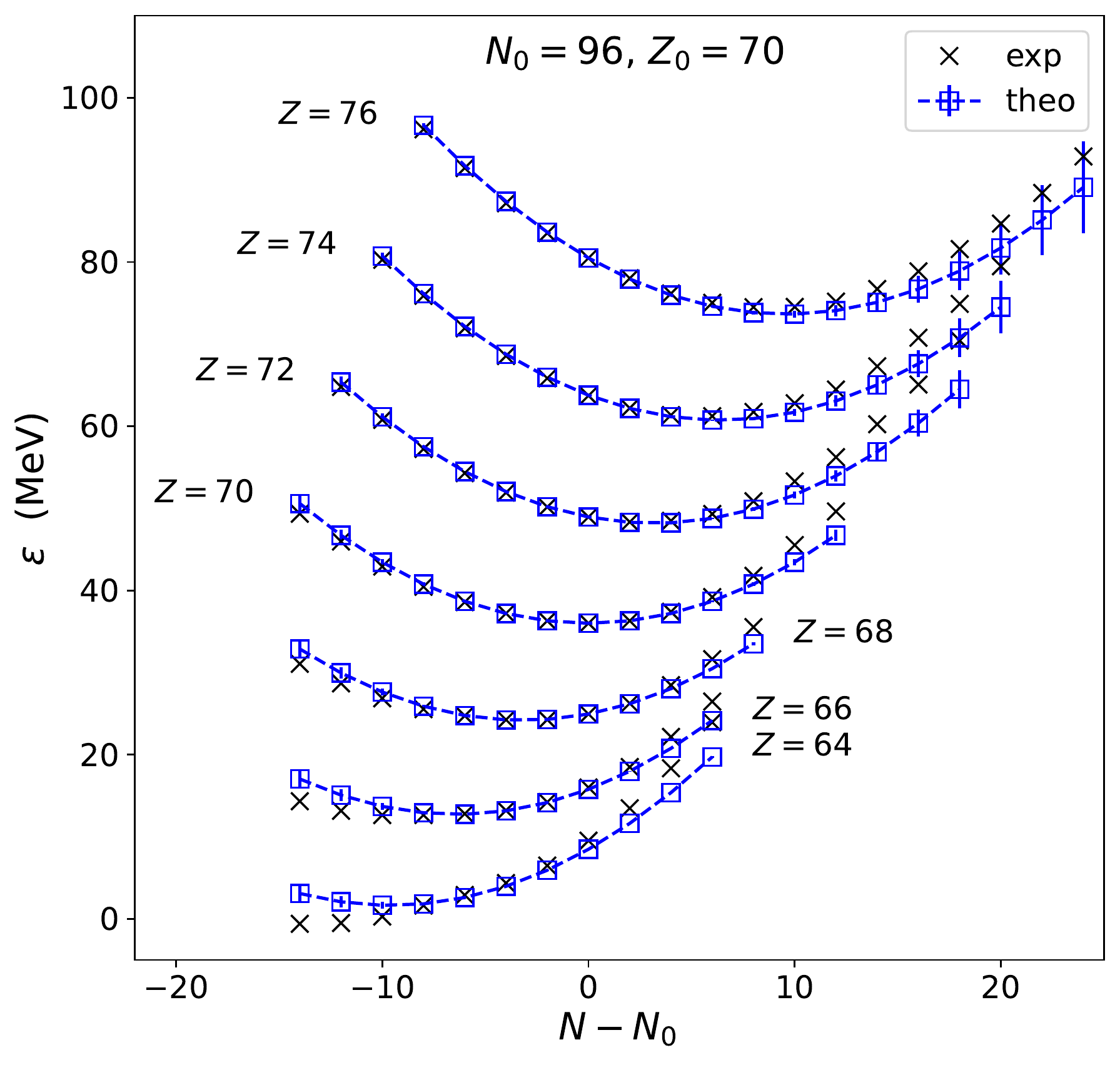}
    \caption{Neutron-pairing rotational bands as sections of a pairing
      elliptical paraboloid for proton numbers as indicated in the
      rare earth region around $^{166}$Yb ($Z_0=70, N_0=96$):
      Experimental data is compared with the theory prediction. A
      total of six low-energy constant has been adjusted for all shown
      bands. Different bands are shifted by multiples of 12~MeV.}
    \label{fig:Yb166-Npairs}
\end{figure}

Figure~\ref{fig:Yb166-Npairs} shows the neutron-pairing rotational
bands (shifted by multiples of 12~MeV) for fixed charge number
$Z$. Also here, theory describes the data fairly well, and deviations
become more pronounced as $|N-N_0|$ or $|Z-Z_0|$ become large.  The
uncertainty estimates $\Delta\varepsilon_{n,z_0}$ from
Eqs.~(\ref{error2}) reflect some of the deviations but are too small
for small $Z$.

\begin{figure}[!htb]
    \includegraphics[width=0.99\linewidth]{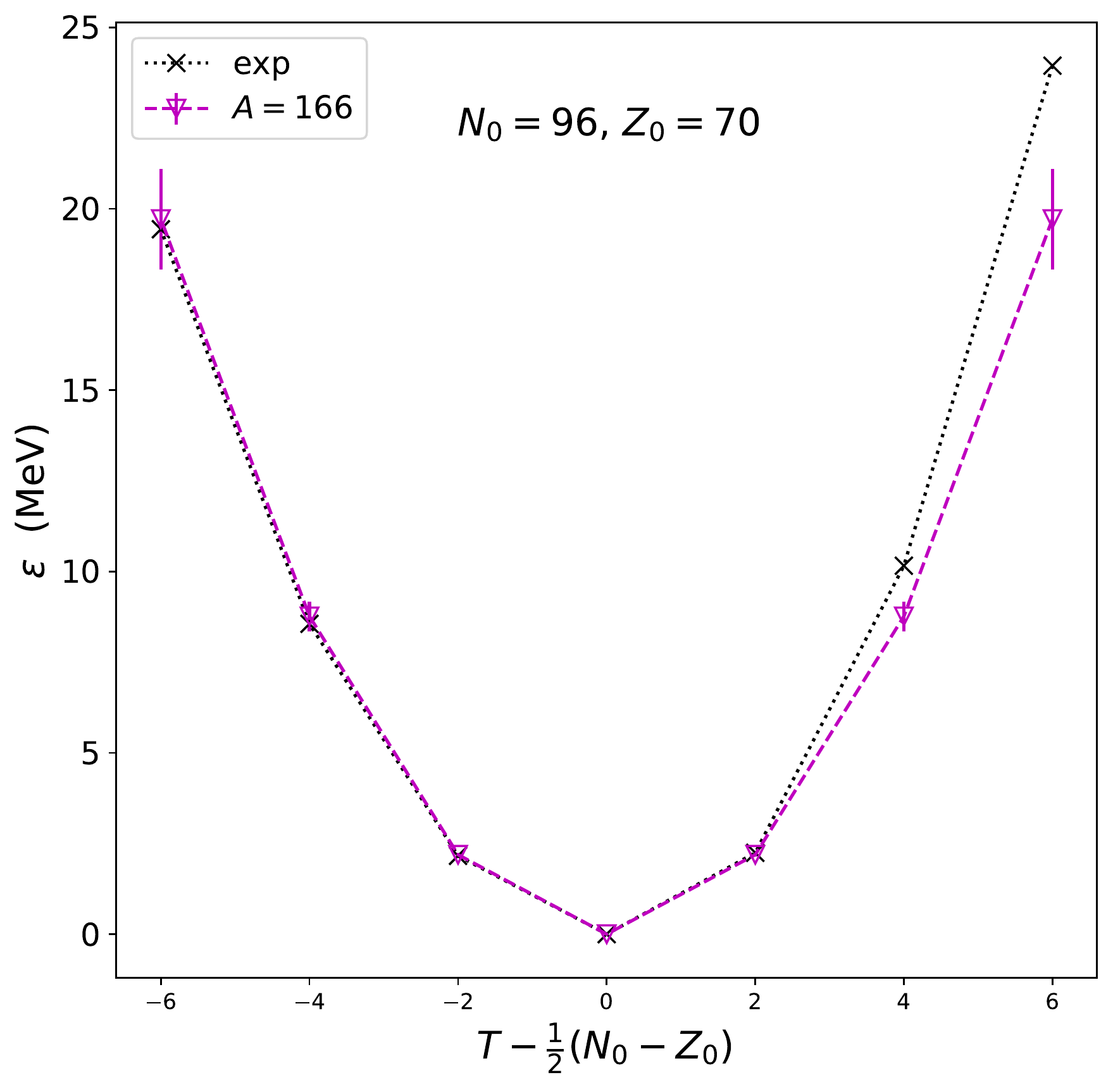}
    \caption{Isobaric rotational band of $A=166$ nuclei as a section
      of the pairing elliptical paraboloid centered at the nucleus
      $^{166}$Yb ($Z_0=70, N_0=96$): Experimental data is compared
      with the theory prediction.}
    \label{fig:Yb-A166}
\end{figure}

The coupling between the two superfluids makes it interesting to also
study other ``directions" of pairing rotational
bands~\cite{hinohara2015}, e.g., the isobar section and the section of
constant isospin projection $T_z$ of the pairing elliptical
paraboloid~(\ref{spec2}).  The former section consists of nuclei that
are connected via double charge exchange reactions, while the latter
section describes nuclei that are linked by $\alpha$ particle capture
or removal. The nucleus $^{166}$Yb is kept at the center.
Figure~\ref{fig:Yb-A166} shows the isobar section. Uncertainty
estimates, taken as $\Delta\varepsilon(T,A_0)$ from
Eqs.~(\ref{error2}), capture the scale of the difference to data but
are not quantitatively correct.

\begin{figure}[!htb]
    \includegraphics[width=0.99\linewidth]{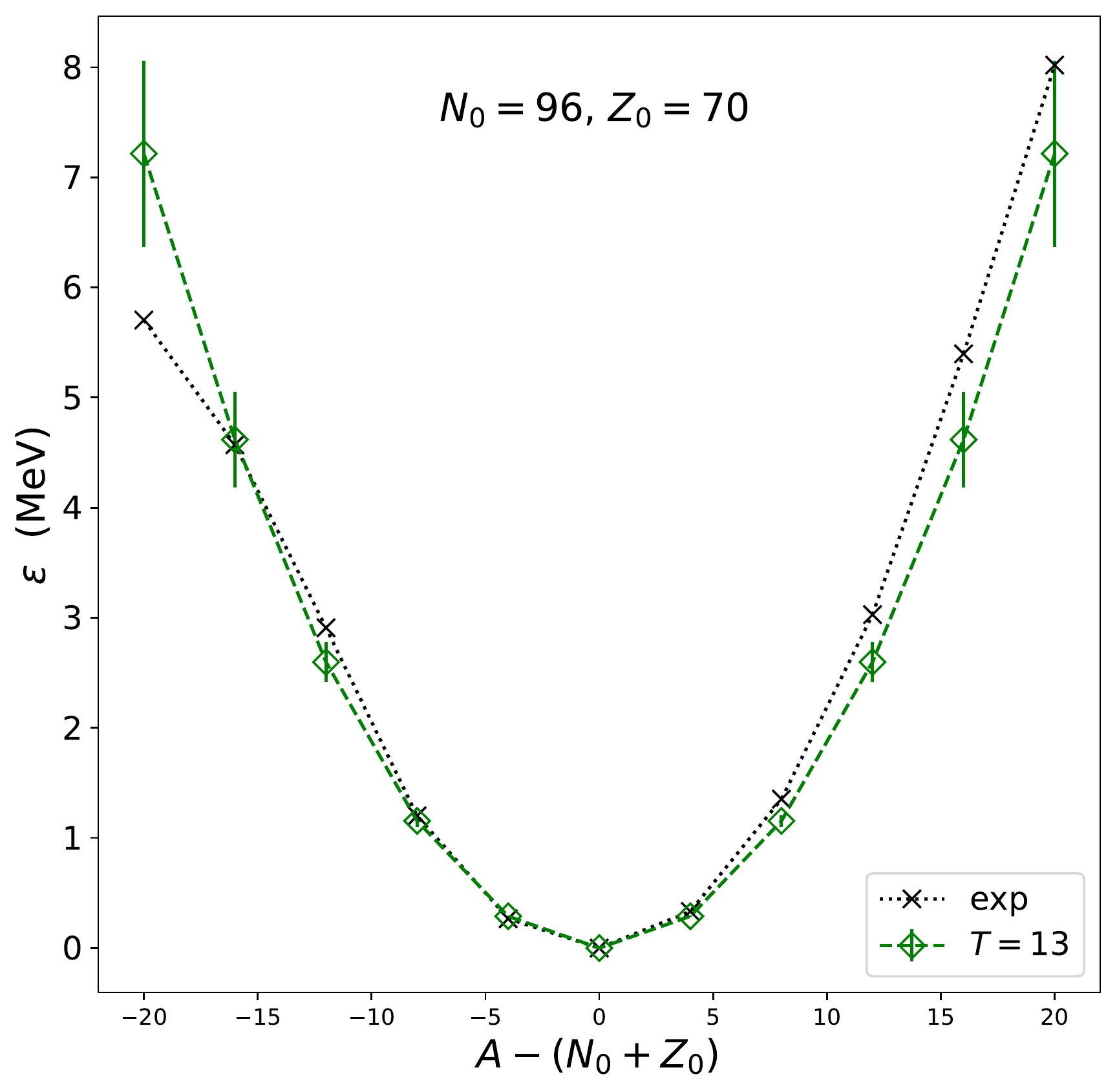}
    \caption{Pairing rotational band with fixed isospin projection
      $T=13$ as a section of the pairing elliptical paraboloid
      centered at the nucleus $^{166}$Yb ($Z_0=70, N_0=96$):
      Experimental data is compared with the theory prediction.}
    \label{fig:Yb-Tz13-alphas}
\end{figure}

Figure~\ref{fig:Yb-Tz13-alphas} shows the section with constant
isospin projection. Here, the uncertainties are taken as
$\Delta\varepsilon(T_0,A)$ from Eqs.~(\ref{error2}). They capture the
scale of differences between theory and data well.

The comparison of the isobar and constant $T_z$ pairing rotational
bands with the $N=96$ proton pairing band of
Fig.~\ref{fig:Yb166-Zpairs} and the $Z=70$ neutron pairing band of
Fig.~\ref{fig:Yb166-Npairs} shows that the rotational constants differ
considerably for each section of the elliptical
paraboloid. Diagonalization of the mass matrix yields eigenvalues
$0.09$ and $2.4$~MeV, and the corresponding eigenvectors have an angle
of $28^\circ$ and $118^\circ$ with the neutron axis on the Segr{\`e}
chart, respectively. (This is essentially along the valley of $\beta$
stability and perpendicular to it.) Consistent with this, the neutron
pairing bands and the constant-$T_z$ pairing band have the smallest
curvature because they are oriented mainly along the valley of $\beta$
stability.

\subsection{Estimating energy gains from particle number projection}
Let us also consider another application of Eq.~(\ref{master2}).
Calculations based on nuclear energy density
functionals~\cite{bender2003,niksic2011} or
Hamiltonians~\cite{dickhoff2004,soma2013} often do not employ particle
number projections.  Then, one really computes a symmetry-breaking
state (with a fixed orientation in gauge space), that consists of a
superposition of states with different numbers of pairs. Such a
localized state clearly has too much kinetic energy in gauge space,
and the formula~(\ref{spec2}) allows one to estimate this. Using
$\langle \hat{N}\rangle = N_0$ and $\Delta N^2\equiv \langle
(\hat{N}-N_0)^2\rangle$, and similar for $\hat{Z}$, one finds
\ba
\label{estimate}
\Delta E&=&{1\over 8a} \langle \Delta N^2\rangle +{1\over 8b}\langle \Delta Z^2\rangle +{1\over 4c}\langle \Delta N \Delta Z\rangle \ .
\ea
Here, the coefficients $a$, $b$, and $c$ may be determined from
computations or data via Eqs.~(\ref{sol2}) and (\ref{solc}).

As an example, let us consider the computation of semi-magic $^{64}$Ni within
Bogoliubov many-body perturbation theory in
Ref.~\cite{tichai2020}. The number variance is about $\Delta N^2\approx16$,
(see Fig.~9 of that work) and $(2a)^{-1}\approx 0.72$~MeV (from
data). This yields $\Delta E\approx 2.9$~MeV.


\section{Summary}
\label{sec:sum}
This paper revisited pairing rotations in a model-independent way
within an effective field theory. It followed the standard approach to
emergent symmetry breaking via a nonlinear realization of the broken
phase symmetry.  This led to pairing rotational bands in semi-magic
nuclei and to a pairing elliptical paraboloid in systems where paired
protons and neutrons interact. Coupling a fermion to the superfluid
extends the theory to odd semi-magic nuclei. The expansion of the
effective Hamiltonians is in powers of differences of Cooper-pair
numbers, and subleading corrections are suppressed by inverse powers
of the maximum number of pairs in a shell.  The key input for the
effective field theory consist of the matrix containing the pairing
rotational constants. The eigenvalues of this model-independent
quantity are given by the curvatures of the nuclear ground-state
energies as a function of proton and neutron numbers.  A comparison
with data shows that the leading-order theory is accurate (within
uncertainty estimates) for heavy semi-magic nuclei and for nuclei
sufficiently far away from shell closures.

The theory predicts that pair transfer is constant for nuclei in a
pairing rotational band. For nuclei on a pairing elliptical
paraboloid, the nuclear matrix element for pair transfer, double
charge exchange reactions, and $\alpha$ particle knockout or capture
are nucleus independent and related to each other.

It is interesting to compare the effective theory of this work with
the those for deformed
nuclei~\cite{papenbrock2011,papenbrock2014,chen2017,papenbrock2020,alnamlah2021}. For
axially symmetric deformations, one exploits the emergent symmetry
breaking of rotational $SO(3)$ down to axial $SO(2)$. Then the coset
spaces is the two-sphere and Nambu-Goldstone modes parameterize that
manifold. Finite ground-state spins and fermions introduce couplings
to gauge potentials (which usually are referred to as Coriolis
forces). The treatment of pairing is technically somewhat simpler than
deformation because the broken symmetry groups are Abelian. Otherwise,
however, one follows the same path.

One could combine both approaches, simultaneously
capturing deformation and superfluidity. Then, the low-energy physics
of nuclei away from shell closures becomes extremely simple: 
The pattern of the emergent symmetry breaking -- from a
product of rotational $SO(3)$ times pairing $U(1) \times U(1)$ down to
axial $SO(2)$ -- is all that matters. The symmetries are realized
nonlinearly, and low-lying excitations are the quantized excitations
of the corresponding Nambu-Goldstone modes in finite systems. Each
nucleus exhibits a ground-state rotational band and pairing rotations
connect ground-state energies of different nuclei.  While we have, of
course, many nuclear models that break symmetries or incorporate the
effects of symmetry breaking, the effective field theory approach makes it
front and center, is aware about its breakdown scale, and allows one
to make systematic improvements and uncertainty estimates.


\begin{acknowledgments}
This work has been supported by the U.S.  Department of Energy under
grant No. DE-FG02-96ER40963 and under contract DE-AC05-00OR22725 with
UT-Battelle, LLC (Oak Ridge National Laboratory).
\end{acknowledgments}


\begin{thebibliography}{51}%
\makeatletter
\providecommand \@ifxundefined [1]{%
 \@ifx{#1\undefined}
}%
\providecommand \@ifnum [1]{%
 \ifnum #1\expandafter \@firstoftwo
 \else \expandafter \@secondoftwo
 \fi
}%
\providecommand \@ifx [1]{%
 \ifx #1\expandafter \@firstoftwo
 \else \expandafter \@secondoftwo
 \fi
}%
\providecommand \natexlab [1]{#1}%
\providecommand \enquote  [1]{``#1''}%
\providecommand \bibnamefont  [1]{#1}%
\providecommand \bibfnamefont [1]{#1}%
\providecommand \citenamefont [1]{#1}%
\providecommand \href@noop [0]{\@secondoftwo}%
\providecommand \href [0]{\begingroup \@sanitize@url \@href}%
\providecommand \@href[1]{\@@startlink{#1}\@@href}%
\providecommand \@@href[1]{\endgroup#1\@@endlink}%
\providecommand \@sanitize@url [0]{\catcode `\\12\catcode `\$12\catcode
  `\&12\catcode `\#12\catcode `\^12\catcode `\_12\catcode `\%12\relax}%
\providecommand \@@startlink[1]{}%
\providecommand \@@endlink[0]{}%
\providecommand \url  [0]{\begingroup\@sanitize@url \@url }%
\providecommand \@url [1]{\endgroup\@href {#1}{\urlprefix }}%
\providecommand \urlprefix  [0]{URL }%
\providecommand \Eprint [0]{\href }%
\providecommand \doibase [0]{http://dx.doi.org/}%
\providecommand \selectlanguage [0]{\@gobble}%
\providecommand \bibinfo  [0]{\@secondoftwo}%
\providecommand \bibfield  [0]{\@secondoftwo}%
\providecommand \translation [1]{[#1]}%
\providecommand \BibitemOpen [0]{}%
\providecommand \bibitemStop [0]{}%
\providecommand \bibitemNoStop [0]{.\EOS\space}%
\providecommand \EOS [0]{\spacefactor3000\relax}%
\providecommand \BibitemShut  [1]{\csname bibitem#1\endcsname}%
\let\auto@bib@innerbib\@empty
\bibitem [{\citenamefont {Bohr}\ \emph {et~al.}(1958)\citenamefont {Bohr},
  \citenamefont {Mottelson},\ and\ \citenamefont {Pines}}]{bohr1958}%
  \BibitemOpen
  \bibfield  {author} {\bibinfo {author} {\bibfnamefont {A.}~\bibnamefont
  {Bohr}}, \bibinfo {author} {\bibfnamefont {B.~R.}\ \bibnamefont {Mottelson}},
  \ and\ \bibinfo {author} {\bibfnamefont {D.}~\bibnamefont {Pines}},\
  }\bibfield  {title} {\enquote {\bibinfo {title} {Possible analogy between the
  excitation spectra of nuclei and those of the superconducting metallic
  state},}\ }\href {\doibase 10.1103/PhysRev.110.936} {\bibfield  {journal}
  {\bibinfo  {journal} {Phys. Rev.}\ }\textbf {\bibinfo {volume} {110}},\
  \bibinfo {pages} {936--938} (\bibinfo {year} {1958})}\BibitemShut {NoStop}%
\bibitem [{\citenamefont {Migdal}(1959)}]{migdal1959}%
  \BibitemOpen
  \bibfield  {author} {\bibinfo {author} {\bibfnamefont {A.~B.}\ \bibnamefont
  {Migdal}},\ }\bibfield  {title} {\enquote {\bibinfo {title} {Superfluidity
  and the moments of inertia of nuclei},}\ }\href {\doibase
  10.1016/0029-5582(59)90264-0} {\bibfield  {journal} {\bibinfo  {journal}
  {Nuclear Physics}\ }\textbf {\bibinfo {volume} {13}},\ \bibinfo {pages}
  {655--674} (\bibinfo {year} {1959})}\BibitemShut {NoStop}%
\bibitem [{\citenamefont {B{\`e}s}\ and\ \citenamefont
  {Broglia}(1966)}]{bes1966}%
  \BibitemOpen
  \bibfield  {author} {\bibinfo {author} {\bibfnamefont {D.~R.}\ \bibnamefont
  {B{\`e}s}}\ and\ \bibinfo {author} {\bibfnamefont {R.~A.}\ \bibnamefont
  {Broglia}},\ }\bibfield  {title} {\enquote {\bibinfo {title} {Pairing
  vibrations},}\ }\href {\doibase 10.1016/0029-5582(66)90090-3} {\bibfield
  {journal} {\bibinfo  {journal} {Nuclear Physics}\ }\textbf {\bibinfo {volume}
  {80}},\ \bibinfo {pages} {289--313} (\bibinfo {year} {1966})}\BibitemShut
  {NoStop}%
\bibitem [{\citenamefont {Nogami}(1964)}]{nogami1964}%
  \BibitemOpen
  \bibfield  {author} {\bibinfo {author} {\bibfnamefont {Yukihisa}\
  \bibnamefont {Nogami}},\ }\bibfield  {title} {\enquote {\bibinfo {title}
  {Improved superconductivity approximation for the pairing interaction in
  nuclei},}\ }\href {\doibase 10.1103/PhysRev.134.B313} {\bibfield  {journal}
  {\bibinfo  {journal} {Phys. Rev.}\ }\textbf {\bibinfo {volume} {134}},\
  \bibinfo {pages} {B313--B321} (\bibinfo {year} {1964})}\BibitemShut {NoStop}%
\bibitem [{\citenamefont {Bohr}(1969)}]{bohr1969}%
  \BibitemOpen
  \bibfield  {author} {\bibinfo {author} {\bibfnamefont {A.}~\bibnamefont
  {Bohr}},\ }\bibfield  {title} {\enquote {\bibinfo {title} {Pair correlations
  and double transfer reactions},}\ }in\ \href
  {https://www.iaea.org/publications/2149/nuclear-structure-dubna-symposium-1968-dubna-4-11-july-1968}
  {\emph {\bibinfo {booktitle} {Nuclear Structure: Dubna Symposium 1968 (Dubna,
  4-11 July 1968)}}},\ \bibinfo {series and number} {Proceedings Series}\
  (\bibinfo  {publisher} {International Atomic Energy Agency},\ \bibinfo
  {address} {Vienna},\ \bibinfo {year} {1969})\ p.\ \bibinfo {pages}
  {179}\BibitemShut {NoStop}%
\bibitem [{\citenamefont {B{\`e}s}\ \emph {et~al.}(1970)\citenamefont
  {B{\`e}s}, \citenamefont {Broglia}, \citenamefont {Perazzo},\ and\
  \citenamefont {Kumar}}]{bes1970}%
  \BibitemOpen
  \bibfield  {author} {\bibinfo {author} {\bibfnamefont {D.~R.}\ \bibnamefont
  {B{\`e}s}}, \bibinfo {author} {\bibfnamefont {R.~A.}\ \bibnamefont
  {Broglia}}, \bibinfo {author} {\bibfnamefont {R.~P.~J.}\ \bibnamefont
  {Perazzo}}, \ and\ \bibinfo {author} {\bibfnamefont {K.}~\bibnamefont
  {Kumar}},\ }\bibfield  {title} {\enquote {\bibinfo {title} {Collective
  treatment of the pairing hamiltonian: (i). formulation of the model},}\
  }\href {\doibase 10.1016/0375-9474(70)90677-9} {\bibfield  {journal}
  {\bibinfo  {journal} {Nuclear Physics A}\ }\textbf {\bibinfo {volume}
  {143}},\ \bibinfo {pages} {1--33} (\bibinfo {year} {1970})}\BibitemShut
  {NoStop}%
\bibitem [{\citenamefont {Broglia}\ \emph {et~al.}(1973)\citenamefont
  {Broglia}, \citenamefont {Hansen},\ and\ \citenamefont
  {Riedel}}]{broglia1973}%
  \BibitemOpen
  \bibfield  {author} {\bibinfo {author} {\bibfnamefont {R.~A.}\ \bibnamefont
  {Broglia}}, \bibinfo {author} {\bibfnamefont {O.}~\bibnamefont {Hansen}}, \
  and\ \bibinfo {author} {\bibfnamefont {C.}~\bibnamefont {Riedel}},\
  }\bibfield  {title} {\enquote {\bibinfo {title} {Two-neutron transfer
  reactions and the pairing model},}\ }in\ \href {\doibase
  10.1007/978-1-4615-9041-5_3} {\emph {\bibinfo {booktitle} {Advances in
  Nuclear Physics}}},\ Vol.~\bibinfo {volume} {6},\ \bibinfo {editor} {edited
  by\ \bibinfo {editor} {\bibfnamefont {M.}~\bibnamefont {Baranger}}\ and\
  \bibinfo {editor} {\bibfnamefont {E.}~\bibnamefont {Vogt}}}\ (\bibinfo
  {publisher} {Springer},\ \bibinfo {address} {Boston, MA},\ \bibinfo {year}
  {1973})\ Chap.~\bibinfo {chapter} {3}, p.\ \bibinfo {pages} {287}\BibitemShut
  {NoStop}%
\bibitem [{\citenamefont {Brink}\ and\ \citenamefont
  {Broglia}(2005)}]{brink_broglia_2005}%
  \BibitemOpen
  \bibfield  {author} {\bibinfo {author} {\bibfnamefont {David~M.}\
  \bibnamefont {Brink}}\ and\ \bibinfo {author} {\bibfnamefont {Ricardo~A.}\
  \bibnamefont {Broglia}},\ }\href {\doibase 10.1017/CBO9780511534911} {\emph
  {\bibinfo {title} {Nuclear Superfluidity: Pairing in Finite Systems}}},\
  Cambridge Monographs on Particle Physics, Nuclear Physics and Cosmology\
  (\bibinfo  {publisher} {Cambridge University Press},\ \bibinfo {year}
  {2005})\BibitemShut {NoStop}%
\bibitem [{\citenamefont {Broglia}\ \emph {et~al.}(2000)\citenamefont
  {Broglia}, \citenamefont {Terasaki},\ and\ \citenamefont
  {Giovanardi}}]{broglia2000}%
  \BibitemOpen
  \bibfield  {author} {\bibinfo {author} {\bibfnamefont {R.~A.}\ \bibnamefont
  {Broglia}}, \bibinfo {author} {\bibfnamefont {J.}~\bibnamefont {Terasaki}}, \
  and\ \bibinfo {author} {\bibfnamefont {N.}~\bibnamefont {Giovanardi}},\
  }\bibfield  {title} {\enquote {\bibinfo {title} {The
  anderson–goldstone–nambu mode in finite and in infinite systems},}\
  }\href {\doibase 10.1016/S0370-1573(00)00046-6} {\bibfield  {journal}
  {\bibinfo  {journal} {Physics Reports}\ }\textbf {\bibinfo {volume} {335}},\
  \bibinfo {pages} {1--18} (\bibinfo {year} {2000})}\BibitemShut {NoStop}%
\bibitem [{\citenamefont {Hinohara}\ and\ \citenamefont
  {Nazarewicz}(2016)}]{hinohara2016}%
  \BibitemOpen
  \bibfield  {author} {\bibinfo {author} {\bibfnamefont {Nobuo}\ \bibnamefont
  {Hinohara}}\ and\ \bibinfo {author} {\bibfnamefont {Witold}\ \bibnamefont
  {Nazarewicz}},\ }\bibfield  {title} {\enquote {\bibinfo {title} {Pairing
  nambu-goldstone modes within nuclear density functional theory},}\ }\href
  {\doibase 10.1103/PhysRevLett.116.152502} {\bibfield  {journal} {\bibinfo
  {journal} {Phys. Rev. Lett.}\ }\textbf {\bibinfo {volume} {116}},\ \bibinfo
  {pages} {152502} (\bibinfo {year} {2016})}\BibitemShut {NoStop}%
\bibitem [{\citenamefont {Potel}\ \emph {et~al.}(2017)\citenamefont {Potel},
  \citenamefont {Idini}, \citenamefont {Barranco}, \citenamefont {Vigezzi},\
  and\ \citenamefont {Broglia}}]{potel2017}%
  \BibitemOpen
  \bibfield  {author} {\bibinfo {author} {\bibfnamefont {G.}~\bibnamefont
  {Potel}}, \bibinfo {author} {\bibfnamefont {A.}~\bibnamefont {Idini}},
  \bibinfo {author} {\bibfnamefont {F.}~\bibnamefont {Barranco}}, \bibinfo
  {author} {\bibfnamefont {E.}~\bibnamefont {Vigezzi}}, \ and\ \bibinfo
  {author} {\bibfnamefont {R.~A.}\ \bibnamefont {Broglia}},\ }\bibfield
  {title} {\enquote {\bibinfo {title} {From bare to renormalized order
  parameter in gauge space: Structure and reactions},}\ }\href {\doibase
  10.1103/PhysRevC.96.034606} {\bibfield  {journal} {\bibinfo  {journal} {Phys.
  Rev. C}\ }\textbf {\bibinfo {volume} {96}},\ \bibinfo {pages} {034606}
  (\bibinfo {year} {2017})}\BibitemShut {NoStop}%
\bibitem [{\citenamefont {von Oertzen}\ and\ \citenamefont
  {Vitturi}(2001)}]{oertzen2001}%
  \BibitemOpen
  \bibfield  {author} {\bibinfo {author} {\bibfnamefont {W.}~\bibnamefont {von
  Oertzen}}\ and\ \bibinfo {author} {\bibfnamefont {A.}~\bibnamefont
  {Vitturi}},\ }\bibfield  {title} {\enquote {\bibinfo {title} {Pairing
  correlations of nucleons and multi-nucleon transfer between heavy nuclei},}\
  }\href {\doibase 10.1088/0034-4885/64/10/202} {\bibfield  {journal} {\bibinfo
   {journal} {Rep. Prog. Phys.}\ }\textbf {\bibinfo {volume} {64}},\ \bibinfo
  {pages} {1247--1337} (\bibinfo {year} {2001})}\BibitemShut {NoStop}%
\bibitem [{\citenamefont {Potel}\ \emph {et~al.}(2011)\citenamefont {Potel},
  \citenamefont {Barranco}, \citenamefont {Marini}, \citenamefont {Idini},
  \citenamefont {Vigezzi},\ and\ \citenamefont {Broglia}}]{potel2011}%
  \BibitemOpen
  \bibfield  {author} {\bibinfo {author} {\bibfnamefont {G.}~\bibnamefont
  {Potel}}, \bibinfo {author} {\bibfnamefont {F.}~\bibnamefont {Barranco}},
  \bibinfo {author} {\bibfnamefont {F.}~\bibnamefont {Marini}}, \bibinfo
  {author} {\bibfnamefont {A.}~\bibnamefont {Idini}}, \bibinfo {author}
  {\bibfnamefont {E.}~\bibnamefont {Vigezzi}}, \ and\ \bibinfo {author}
  {\bibfnamefont {R.~A.}\ \bibnamefont {Broglia}},\ }\bibfield  {title}
  {\enquote {\bibinfo {title} {Calculation of the transition from pairing
  vibrational to pairing rotational regimes between magic nuclei
  $^{100}\mathrm{Sn}$ and $^{132}\mathrm{Sn}$ via two-nucleon transfer
  reactions},}\ }\href {\doibase 10.1103/PhysRevLett.107.092501} {\bibfield
  {journal} {\bibinfo  {journal} {Phys. Rev. Lett.}\ }\textbf {\bibinfo
  {volume} {107}},\ \bibinfo {pages} {092501} (\bibinfo {year}
  {2011})}\BibitemShut {NoStop}%
\bibitem [{\citenamefont {Potel}\ \emph
  {et~al.}(2013{\natexlab{a}})\citenamefont {Potel}, \citenamefont {Idini},
  \citenamefont {Barranco}, \citenamefont {Vigezzi},\ and\ \citenamefont
  {Broglia}}]{potel2013}%
  \BibitemOpen
  \bibfield  {author} {\bibinfo {author} {\bibfnamefont {G.}~\bibnamefont
  {Potel}}, \bibinfo {author} {\bibfnamefont {A.}~\bibnamefont {Idini}},
  \bibinfo {author} {\bibfnamefont {F.}~\bibnamefont {Barranco}}, \bibinfo
  {author} {\bibfnamefont {E.}~\bibnamefont {Vigezzi}}, \ and\ \bibinfo
  {author} {\bibfnamefont {R.~A.}\ \bibnamefont {Broglia}},\ }\bibfield
  {title} {\enquote {\bibinfo {title} {Cooper pair transfer in nuclei},}\
  }\href {\doibase 10.1088/0034-4885/76/10/106301} {\bibfield  {journal}
  {\bibinfo  {journal} {Rep. Prog. Phys.}\ }\textbf {\bibinfo {volume} {76}},\
  \bibinfo {pages} {106301} (\bibinfo {year} {2013}{\natexlab{a}})}\BibitemShut
  {NoStop}%
\bibitem [{\citenamefont {Potel}\ \emph
  {et~al.}(2013{\natexlab{b}})\citenamefont {Potel}, \citenamefont {Idini},
  \citenamefont {Barranco}, \citenamefont {Vigezzi},\ and\ \citenamefont
  {Broglia}}]{potel2013b}%
  \BibitemOpen
  \bibfield  {author} {\bibinfo {author} {\bibfnamefont {G.}~\bibnamefont
  {Potel}}, \bibinfo {author} {\bibfnamefont {A.}~\bibnamefont {Idini}},
  \bibinfo {author} {\bibfnamefont {F.}~\bibnamefont {Barranco}}, \bibinfo
  {author} {\bibfnamefont {E.}~\bibnamefont {Vigezzi}}, \ and\ \bibinfo
  {author} {\bibfnamefont {R.~A.}\ \bibnamefont {Broglia}},\ }\bibfield
  {title} {\enquote {\bibinfo {title} {Quantitative study of coherent pairing
  modes with two-neutron transfer: Sn isotopes},}\ }\href {\doibase
  10.1103/PhysRevC.87.054321} {\bibfield  {journal} {\bibinfo  {journal} {Phys.
  Rev. C}\ }\textbf {\bibinfo {volume} {87}},\ \bibinfo {pages} {054321}
  (\bibinfo {year} {2013}{\natexlab{b}})}\BibitemShut {NoStop}%
\bibitem [{\citenamefont {Shimoyama}\ and\ \citenamefont
  {Matsuo}(2011)}]{shimoyama2011}%
  \BibitemOpen
  \bibfield  {author} {\bibinfo {author} {\bibfnamefont {Hirotaka}\
  \bibnamefont {Shimoyama}}\ and\ \bibinfo {author} {\bibfnamefont {Masayuki}\
  \bibnamefont {Matsuo}},\ }\bibfield  {title} {\enquote {\bibinfo {title}
  {Anomalous pairing vibration in neutron-rich sn isotopes beyond the $n=82$
  magic number},}\ }\href {\doibase 10.1103/PhysRevC.84.044317} {\bibfield
  {journal} {\bibinfo  {journal} {Phys. Rev. C}\ }\textbf {\bibinfo {volume}
  {84}},\ \bibinfo {pages} {044317} (\bibinfo {year} {2011})}\BibitemShut
  {NoStop}%
\bibitem [{\citenamefont {{Beck}}\ \emph {et~al.}(1972)\citenamefont {{Beck}},
  \citenamefont {{Kleber}},\ and\ \citenamefont {{Schmidt}}}]{beck1972}%
  \BibitemOpen
  \bibfield  {author} {\bibinfo {author} {\bibfnamefont {Rainer}\ \bibnamefont
  {{Beck}}}, \bibinfo {author} {\bibfnamefont {Manfred}\ \bibnamefont
  {{Kleber}}}, \ and\ \bibinfo {author} {\bibfnamefont {Hartwig}\ \bibnamefont
  {{Schmidt}}},\ }\bibfield  {title} {\enquote {\bibinfo {title} {{Pairing
  rotations and separation energies}},}\ }\href {\doibase 10.1007/BF01386946}
  {\bibfield  {journal} {\bibinfo  {journal} {Zeitschrift fur Physik}\ }\textbf
  {\bibinfo {volume} {250}},\ \bibinfo {pages} {155--165} (\bibinfo {year}
  {1972})}\BibitemShut {NoStop}%
\bibitem [{\citenamefont {Matsuo}(1986)}]{matsuo1986b}%
  \BibitemOpen
  \bibfield  {author} {\bibinfo {author} {\bibfnamefont {Masayuki}\
  \bibnamefont {Matsuo}},\ }\bibfield  {title} {\enquote {\bibinfo {title}
  {{Treatment of Nucleon-Number Conservation in the Selfconsistent
  Collective-Coordinate Method: -Coupling between Large-Amplitude Collective
  Motion and Pairing Rotation—}},}\ }\href {\doibase 10.1143/PTP.76.372}
  {\bibfield  {journal} {\bibinfo  {journal} {Progress of Theoretical Physics}\
  }\textbf {\bibinfo {volume} {76}},\ \bibinfo {pages} {372--386} (\bibinfo
  {year} {1986})}\BibitemShut {NoStop}%
\bibitem [{\citenamefont {Hinohara}(2015)}]{hinohara2015}%
  \BibitemOpen
  \bibfield  {author} {\bibinfo {author} {\bibfnamefont {Nobuo}\ \bibnamefont
  {Hinohara}},\ }\bibfield  {title} {\enquote {\bibinfo {title} {Collective
  inertia of the nambu-goldstone mode from linear response theory},}\ }\href
  {\doibase 10.1103/PhysRevC.92.034321} {\bibfield  {journal} {\bibinfo
  {journal} {Phys. Rev. C}\ }\textbf {\bibinfo {volume} {92}},\ \bibinfo
  {pages} {034321} (\bibinfo {year} {2015})}\BibitemShut {NoStop}%
\bibitem [{\citenamefont {Hinohara}(2018)}]{hinohara2018}%
  \BibitemOpen
  \bibfield  {author} {\bibinfo {author} {\bibfnamefont {Nobuo}\ \bibnamefont
  {Hinohara}},\ }\bibfield  {title} {\enquote {\bibinfo {title} {Extending
  pairing energy density functional using pairing rotational moments of
  inertia},}\ }\href {\doibase 10.1088/1361-6471/aa9f8b} {\bibfield  {journal}
  {\bibinfo  {journal} {J. Phys. G: Nucl. Part. Phys.}\ }\textbf {\bibinfo
  {volume} {45}},\ \bibinfo {pages} {024004} (\bibinfo {year}
  {2018})}\BibitemShut {NoStop}%
\bibitem [{\citenamefont {Kouno}\ \emph {et~al.}(2021)\citenamefont {Kouno},
  \citenamefont {Ishizuka}, \citenamefont {Inakura},\ and\ \citenamefont
  {Chiba}}]{kouno2021}%
  \BibitemOpen
  \bibfield  {author} {\bibinfo {author} {\bibfnamefont {Taiki}\ \bibnamefont
  {Kouno}}, \bibinfo {author} {\bibfnamefont {Chikako}\ \bibnamefont
  {Ishizuka}}, \bibinfo {author} {\bibfnamefont {Tsunenori}\ \bibnamefont
  {Inakura}}, \ and\ \bibinfo {author} {\bibfnamefont {Satoshi}\ \bibnamefont
  {Chiba}},\ }\bibfield  {title} {\enquote {\bibinfo {title} {{Pairing strength
  in the relativistic mean-field theory determined from the fission barrier
  heights of actinide nuclei and verified by pairing rotation and binding
  energies}},}\ }\href {\doibase 10.1093/ptep/ptab167} {\bibfield  {journal}
  {\bibinfo  {journal} {Progress of Theoretical and Experimental Physics}\
  }\textbf {\bibinfo {volume} {2022}} (\bibinfo {year} {2021}),\
  10.1093/ptep/ptab167},\ \bibinfo {note} {023D02}\BibitemShut {NoStop}%
\bibitem [{\citenamefont {Kolck}(1999)}]{vankolck1999}%
  \BibitemOpen
  \bibfield  {author} {\bibinfo {author} {\bibfnamefont {U.~Van}\ \bibnamefont
  {Kolck}},\ }\bibfield  {title} {\enquote {\bibinfo {title} {Effective field
  theory of nuclear forces},}\ }\href {\doibase 10.1016/S0146-6410(99)00097-6}
  {\bibfield  {journal} {\bibinfo  {journal} {Prog. Part. Nucl. Phys.}\
  }\textbf {\bibinfo {volume} {43}},\ \bibinfo {pages} {337 -- 418} (\bibinfo
  {year} {1999})}\BibitemShut {NoStop}%
\bibitem [{\citenamefont {Hammer}\ and\ \citenamefont
  {Furnstahl}(2000)}]{hammer2000}%
  \BibitemOpen
  \bibfield  {author} {\bibinfo {author} {\bibfnamefont {H.-W.}\ \bibnamefont
  {Hammer}}\ and\ \bibinfo {author} {\bibfnamefont {R.J.}\ \bibnamefont
  {Furnstahl}},\ }\bibfield  {title} {\enquote {\bibinfo {title} {Effective
  field theory for dilute fermi systems},}\ }\href {\doibase
  10.1016/S0375-9474(00)00325-0} {\bibfield  {journal} {\bibinfo  {journal}
  {Nuclear Physics A}\ }\textbf {\bibinfo {volume} {678}},\ \bibinfo {pages}
  {277 -- 294} (\bibinfo {year} {2000})}\BibitemShut {NoStop}%
\bibitem [{\citenamefont {{Bedaque}}\ and\ \citenamefont {{van
  Kolck}}(2002)}]{bedaque2002}%
  \BibitemOpen
  \bibfield  {author} {\bibinfo {author} {\bibfnamefont {P.~F.}\ \bibnamefont
  {{Bedaque}}}\ and\ \bibinfo {author} {\bibfnamefont {U.}~\bibnamefont {{van
  Kolck}}},\ }\bibfield  {title} {\enquote {\bibinfo {title} {{Effective field
  theory for few-nucleon systems}},}\ }\href {\doibase
  10.1146/annurev.nucl.52.050102.090637} {\bibfield  {journal} {\bibinfo
  {journal} {Annual Review of Nuclear and Particle Science}\ }\textbf {\bibinfo
  {volume} {52}},\ \bibinfo {pages} {339--396} (\bibinfo {year} {2002})},\
  \Eprint {http://arxiv.org/abs/nucl-th/0203055} {nucl-th/0203055} \BibitemShut
  {NoStop}%
\bibitem [{\citenamefont {Furnstahl}\ \emph {et~al.}(2007)\citenamefont
  {Furnstahl}, \citenamefont {Hammer},\ and\ \citenamefont
  {Puglia}}]{furnstahl2007}%
  \BibitemOpen
  \bibfield  {author} {\bibinfo {author} {\bibfnamefont {R.J.}\ \bibnamefont
  {Furnstahl}}, \bibinfo {author} {\bibfnamefont {H.-W.}\ \bibnamefont
  {Hammer}}, \ and\ \bibinfo {author} {\bibfnamefont {S.J.}\ \bibnamefont
  {Puglia}},\ }\bibfield  {title} {\enquote {\bibinfo {title} {Effective field
  theory for dilute fermions with pairing},}\ }\href {\doibase
  10.1016/j.aop.2007.01.003} {\bibfield  {journal} {\bibinfo  {journal} {Annals
  of Physics}\ }\textbf {\bibinfo {volume} {322}},\ \bibinfo {pages} {2703 --
  2732} (\bibinfo {year} {2007})}\BibitemShut {NoStop}%
\bibitem [{\citenamefont {Epelbaum}\ \emph {et~al.}(2009)\citenamefont
  {Epelbaum}, \citenamefont {Hammer},\ and\ \citenamefont
  {Mei\ss{}ner}}]{epelbaum2009}%
  \BibitemOpen
  \bibfield  {author} {\bibinfo {author} {\bibfnamefont {E.}~\bibnamefont
  {Epelbaum}}, \bibinfo {author} {\bibfnamefont {H.-W.}\ \bibnamefont
  {Hammer}}, \ and\ \bibinfo {author} {\bibfnamefont {Ulf-G.}\ \bibnamefont
  {Mei\ss{}ner}},\ }\bibfield  {title} {\enquote {\bibinfo {title} {Modern
  theory of nuclear forces},}\ }\href {\doibase 10.1103/RevModPhys.81.1773}
  {\bibfield  {journal} {\bibinfo  {journal} {Rev. Mod. Phys.}\ }\textbf
  {\bibinfo {volume} {81}},\ \bibinfo {pages} {1773--1825} (\bibinfo {year}
  {2009})}\BibitemShut {NoStop}%
\bibitem [{\citenamefont {Papenbrock}(2011)}]{papenbrock2011}%
  \BibitemOpen
  \bibfield  {author} {\bibinfo {author} {\bibfnamefont {T.}~\bibnamefont
  {Papenbrock}},\ }\bibfield  {title} {\enquote {\bibinfo {title} {Effective
  theory for deformed nuclei},}\ }\href {\doibase
  10.1016/j.nuclphysa.2010.12.013} {\bibfield  {journal} {\bibinfo  {journal}
  {Nucl. Phys. A}\ }\textbf {\bibinfo {volume} {852}},\ \bibinfo {pages} {36 --
  60} (\bibinfo {year} {2011})}\BibitemShut {NoStop}%
\bibitem [{\citenamefont {Grie{\ss}hammer}\ \emph {et~al.}(2012)\citenamefont
  {Grie{\ss}hammer}, \citenamefont {McGovern}, \citenamefont {Phillips},\ and\
  \citenamefont {Feldman}}]{griesshammer2012}%
  \BibitemOpen
  \bibfield  {author} {\bibinfo {author} {\bibfnamefont {H.~W.}\ \bibnamefont
  {Grie{\ss}hammer}}, \bibinfo {author} {\bibfnamefont {J.~A.}\ \bibnamefont
  {McGovern}}, \bibinfo {author} {\bibfnamefont {D.~R.}\ \bibnamefont
  {Phillips}}, \ and\ \bibinfo {author} {\bibfnamefont {G.}~\bibnamefont
  {Feldman}},\ }\bibfield  {title} {\enquote {\bibinfo {title} {Using effective
  field theory to analyse low-energy compton scattering data from protons and
  light nuclei},}\ }\href {\doibase 10.1016/j.ppnp.2012.04.003} {\bibfield
  {journal} {\bibinfo  {journal} {Prog. Part. Nucl. Phys.}\ }\textbf {\bibinfo
  {volume} {67}},\ \bibinfo {pages} {841 -- 897} (\bibinfo {year}
  {2012})}\BibitemShut {NoStop}%
\bibitem [{\citenamefont {Hammer}\ \emph {et~al.}(2017)\citenamefont {Hammer},
  \citenamefont {Ji},\ and\ \citenamefont {Phillips}}]{hammer2017}%
  \BibitemOpen
  \bibfield  {author} {\bibinfo {author} {\bibfnamefont {H.-W.}\ \bibnamefont
  {Hammer}}, \bibinfo {author} {\bibfnamefont {C.}~\bibnamefont {Ji}}, \ and\
  \bibinfo {author} {\bibfnamefont {D.~R.}\ \bibnamefont {Phillips}},\
  }\bibfield  {title} {\enquote {\bibinfo {title} {Effective field theory
  description of halo nuclei},}\ }\href {\doibase 10.1088/1361-6471/aa83db}
  {\bibfield  {journal} {\bibinfo  {journal} {Journal of Physics G: Nuclear and
  Particle Physics}\ }\textbf {\bibinfo {volume} {44}},\ \bibinfo {pages}
  {103002} (\bibinfo {year} {2017})}\BibitemShut {NoStop}%
\bibitem [{\citenamefont {Hammer}\ \emph {et~al.}(2020)\citenamefont {Hammer},
  \citenamefont {K\"onig},\ and\ \citenamefont {van Kolck}}]{hammer2020}%
  \BibitemOpen
  \bibfield  {author} {\bibinfo {author} {\bibfnamefont {H.-W.}\ \bibnamefont
  {Hammer}}, \bibinfo {author} {\bibfnamefont {Sebastian}\ \bibnamefont
  {K\"onig}}, \ and\ \bibinfo {author} {\bibfnamefont {U.}~\bibnamefont {van
  Kolck}},\ }\bibfield  {title} {\enquote {\bibinfo {title} {Nuclear effective
  field theory: Status and perspectives},}\ }\href {\doibase
  10.1103/RevModPhys.92.025004} {\bibfield  {journal} {\bibinfo  {journal}
  {Rev. Mod. Phys.}\ }\textbf {\bibinfo {volume} {92}},\ \bibinfo {pages}
  {025004} (\bibinfo {year} {2020})}\BibitemShut {NoStop}%
\bibitem [{\citenamefont {Schindler}\ and\ \citenamefont
  {Phillips}(2009)}]{schindler2009}%
  \BibitemOpen
  \bibfield  {author} {\bibinfo {author} {\bibfnamefont {M.~R.}\ \bibnamefont
  {Schindler}}\ and\ \bibinfo {author} {\bibfnamefont {D.~R.}\ \bibnamefont
  {Phillips}},\ }\bibfield  {title} {\enquote {\bibinfo {title} {Bayesian
  methods for parameter estimation in effective field theories},}\ }\href
  {\doibase 10.1016/j.aop.2008.09.003} {\bibfield  {journal} {\bibinfo
  {journal} {Ann. Phys.}\ }\textbf {\bibinfo {volume} {324}},\ \bibinfo {pages}
  {682 -- 708} (\bibinfo {year} {2009})}\BibitemShut {NoStop}%
\bibitem [{\citenamefont {{Furnstahl}}\ \emph {et~al.}(2015)\citenamefont
  {{Furnstahl}}, \citenamefont {{Phillips}},\ and\ \citenamefont
  {{Wesolowski}}}]{furnstahl2014c}%
  \BibitemOpen
  \bibfield  {author} {\bibinfo {author} {\bibfnamefont {R.~J.}\ \bibnamefont
  {{Furnstahl}}}, \bibinfo {author} {\bibfnamefont {D.~R.}\ \bibnamefont
  {{Phillips}}}, \ and\ \bibinfo {author} {\bibfnamefont {S.}~\bibnamefont
  {{Wesolowski}}},\ }\bibfield  {title} {\enquote {\bibinfo {title} {A recipe
  for eft uncertainty quantification in nuclear physics},}\ }\href
  {http://stacks.iop.org/0954-3899/42/i=3/a=034028} {\bibfield  {journal}
  {\bibinfo  {journal} {Journal of Physics G: Nuclear and Particle Physics}\
  }\textbf {\bibinfo {volume} {42}},\ \bibinfo {pages} {034028} (\bibinfo
  {year} {2015})}\BibitemShut {NoStop}%
\bibitem [{\citenamefont {Gasser}\ and\ \citenamefont
  {Leutwyler}(1988)}]{gasser1988}%
  \BibitemOpen
  \bibfield  {author} {\bibinfo {author} {\bibfnamefont {J.}~\bibnamefont
  {Gasser}}\ and\ \bibinfo {author} {\bibfnamefont {H.}~\bibnamefont
  {Leutwyler}},\ }\bibfield  {title} {\enquote {\bibinfo {title} {Spontaneously
  broken symmetries: Effective lagrangians at finite volume},}\ }\href
  {\doibase 10.1016/0550-3213(88)90107-1} {\bibfield  {journal} {\bibinfo
  {journal} {Nuclear Physics B}\ }\textbf {\bibinfo {volume} {307}},\ \bibinfo
  {pages} {763 -- 778} (\bibinfo {year} {1988})}\BibitemShut {NoStop}%
\bibitem [{\citenamefont {Yannouleas}\ and\ \citenamefont
  {Landman}(2007)}]{yannouleas2007}%
  \BibitemOpen
  \bibfield  {author} {\bibinfo {author} {\bibfnamefont {C.}~\bibnamefont
  {Yannouleas}}\ and\ \bibinfo {author} {\bibfnamefont {U.}~\bibnamefont
  {Landman}},\ }\bibfield  {title} {\enquote {\bibinfo {title} {Symmetry
  breaking and quantum correlations in finite systems: studies of quantum dots
  and ultracold bose gases and related nuclear and chemical methods},}\ }\href
  {\doibase 10.1088/0034-4885/70/12/R02} {\bibfield  {journal} {\bibinfo
  {journal} {Rep. Prog. Phys.}\ }\textbf {\bibinfo {volume} {70}},\ \bibinfo
  {pages} {2067} (\bibinfo {year} {2007})}\BibitemShut {NoStop}%
\bibitem [{\citenamefont {Weinberg}(1968)}]{weinberg1968}%
  \BibitemOpen
  \bibfield  {author} {\bibinfo {author} {\bibfnamefont {Steven}\ \bibnamefont
  {Weinberg}},\ }\bibfield  {title} {\enquote {\bibinfo {title} {Nonlinear
  realizations of chiral symmetry},}\ }\href {\doibase
  10.1103/PhysRev.166.1568} {\bibfield  {journal} {\bibinfo  {journal} {Phys.
  Rev.}\ }\textbf {\bibinfo {volume} {166}},\ \bibinfo {pages} {1568--1577}
  (\bibinfo {year} {1968})}\BibitemShut {NoStop}%
\bibitem [{\citenamefont {Callan}\ \emph {et~al.}(1969)\citenamefont {Callan},
  \citenamefont {Coleman}, \citenamefont {Wess},\ and\ \citenamefont
  {Zumino}}]{callan1969}%
  \BibitemOpen
  \bibfield  {author} {\bibinfo {author} {\bibfnamefont {Curtis~G.}\
  \bibnamefont {Callan}}, \bibinfo {author} {\bibfnamefont {Sidney}\
  \bibnamefont {Coleman}}, \bibinfo {author} {\bibfnamefont {J.}~\bibnamefont
  {Wess}}, \ and\ \bibinfo {author} {\bibfnamefont {Bruno}\ \bibnamefont
  {Zumino}},\ }\bibfield  {title} {\enquote {\bibinfo {title} {Structure of
  phenomenological lagrangians. ii},}\ }\href {\doibase
  10.1103/PhysRev.177.2247} {\bibfield  {journal} {\bibinfo  {journal} {Phys.
  Rev.}\ }\textbf {\bibinfo {volume} {177}},\ \bibinfo {pages} {2247--2250}
  (\bibinfo {year} {1969})}\BibitemShut {NoStop}%
\bibitem [{\citenamefont {Coleman}\ \emph {et~al.}(1969)\citenamefont
  {Coleman}, \citenamefont {Wess},\ and\ \citenamefont {Zumino}}]{coleman1969}%
  \BibitemOpen
  \bibfield  {author} {\bibinfo {author} {\bibfnamefont {S.}~\bibnamefont
  {Coleman}}, \bibinfo {author} {\bibfnamefont {J.}~\bibnamefont {Wess}}, \
  and\ \bibinfo {author} {\bibfnamefont {Bruno}\ \bibnamefont {Zumino}},\
  }\bibfield  {title} {\enquote {\bibinfo {title} {Structure of
  phenomenological lagrangians. i},}\ }\href {\doibase
  10.1103/PhysRev.177.2239} {\bibfield  {journal} {\bibinfo  {journal} {Phys.
  Rev.}\ }\textbf {\bibinfo {volume} {177}},\ \bibinfo {pages} {2239--2247}
  (\bibinfo {year} {1969})}\BibitemShut {NoStop}%
\bibitem [{\citenamefont {Brauner}(2010)}]{brauner2010}%
  \BibitemOpen
  \bibfield  {author} {\bibinfo {author} {\bibfnamefont {T.}~\bibnamefont
  {Brauner}},\ }\bibfield  {title} {\enquote {\bibinfo {title} {Spontaneous
  symmetry breaking and nambu-goldstone bosons in quantum many-body systems},}\
  }\href {\doibase 10.3390/sym2020609} {\bibfield  {journal} {\bibinfo
  {journal} {Symmetry}\ }\textbf {\bibinfo {volume} {2}},\ \bibinfo {pages}
  {609--657} (\bibinfo {year} {2010})},\ \Eprint
  {http://arxiv.org/abs/1001.5212} {arXiv:1001.5212} \BibitemShut {NoStop}%
\bibitem [{\citenamefont {Papenbrock}\ and\ \citenamefont
  {Weidenm\"uller}(2014)}]{papenbrock2014}%
  \BibitemOpen
  \bibfield  {author} {\bibinfo {author} {\bibfnamefont {T.}~\bibnamefont
  {Papenbrock}}\ and\ \bibinfo {author} {\bibfnamefont {H.~A.}\ \bibnamefont
  {Weidenm\"uller}},\ }\bibfield  {title} {\enquote {\bibinfo {title}
  {Effective field theory for finite systems with spontaneously broken
  symmetry},}\ }\href {\doibase 10.1103/PhysRevC.89.014334} {\bibfield
  {journal} {\bibinfo  {journal} {Phys. Rev. C}\ }\textbf {\bibinfo {volume}
  {89}},\ \bibinfo {pages} {014334} (\bibinfo {year} {2014})}\BibitemShut
  {NoStop}%
\bibitem [{\citenamefont {{Krappe}}(1975)}]{krappe1975}%
  \BibitemOpen
  \bibfield  {author} {\bibinfo {author} {\bibfnamefont {H.~J.}\ \bibnamefont
  {{Krappe}}},\ }\bibfield  {title} {\enquote {\bibinfo {title} {{On the use of
  a variable moment of pairing}},}\ }\href {\doibase 10.1007/BF01409299}
  {\bibfield  {journal} {\bibinfo  {journal} {Zeitschrift fur Physik A Hadrons
  and Nuclei}\ }\textbf {\bibinfo {volume} {275}},\ \bibinfo {pages} {297--304}
  (\bibinfo {year} {1975})}\BibitemShut {NoStop}%
\bibitem [{\citenamefont {Kishimoto}\ and\ \citenamefont
  {Kammuri}(1985)}]{kishimoto1985}%
  \BibitemOpen
  \bibfield  {author} {\bibinfo {author} {\bibfnamefont {Teruo}\ \bibnamefont
  {Kishimoto}}\ and\ \bibinfo {author} {\bibfnamefont {Tetsuo}\ \bibnamefont
  {Kammuri}},\ }\bibfield  {title} {\enquote {\bibinfo {title} {{Pair Rotation
  in the Dynamical Nuclear Field Theory}},}\ }\href {\doibase
  10.1143/PTP.74.1245} {\bibfield  {journal} {\bibinfo  {journal} {Progress of
  Theoretical Physics}\ }\textbf {\bibinfo {volume} {74}},\ \bibinfo {pages}
  {1245--1263} (\bibinfo {year} {1985})}\BibitemShut {NoStop}%
\bibitem [{\citenamefont {Marshalek}(1977)}]{marshalek1977}%
  \BibitemOpen
  \bibfield  {author} {\bibinfo {author} {\bibfnamefont {E.R.}\ \bibnamefont
  {Marshalek}},\ }\bibfield  {title} {\enquote {\bibinfo {title} {The rpa at
  high spin and conservation laws},}\ }\href {\doibase
  10.1016/0375-9474(77)90461-4} {\bibfield  {journal} {\bibinfo  {journal}
  {Nuclear Physics A}\ }\textbf {\bibinfo {volume} {275}},\ \bibinfo {pages}
  {416--444} (\bibinfo {year} {1977})}\BibitemShut {NoStop}%
\bibitem [{\citenamefont {Wang}\ \emph {et~al.}(2014)\citenamefont {Wang},
  \citenamefont {Dobaczewski}, \citenamefont {Kortelainen}, \citenamefont
  {Yu},\ and\ \citenamefont {Stoitsov}}]{wang2014}%
  \BibitemOpen
  \bibfield  {author} {\bibinfo {author} {\bibfnamefont {X.~B.}\ \bibnamefont
  {Wang}}, \bibinfo {author} {\bibfnamefont {J.}~\bibnamefont {Dobaczewski}},
  \bibinfo {author} {\bibfnamefont {M.}~\bibnamefont {Kortelainen}}, \bibinfo
  {author} {\bibfnamefont {L.~F.}\ \bibnamefont {Yu}}, \ and\ \bibinfo {author}
  {\bibfnamefont {M.~V.}\ \bibnamefont {Stoitsov}},\ }\bibfield  {title}
  {\enquote {\bibinfo {title} {Lipkin method of particle-number restoration to
  higher orders},}\ }\href {\doibase 10.1103/PhysRevC.90.014312} {\bibfield
  {journal} {\bibinfo  {journal} {Phys. Rev. C}\ }\textbf {\bibinfo {volume}
  {90}},\ \bibinfo {pages} {014312} (\bibinfo {year} {2014})}\BibitemShut
  {NoStop}%
\bibitem [{\citenamefont {Bender}\ \emph {et~al.}(2003)\citenamefont {Bender},
  \citenamefont {Heenen},\ and\ \citenamefont {Reinhard}}]{bender2003}%
  \BibitemOpen
  \bibfield  {author} {\bibinfo {author} {\bibfnamefont {Michael}\ \bibnamefont
  {Bender}}, \bibinfo {author} {\bibfnamefont {Paul-Henri}\ \bibnamefont
  {Heenen}}, \ and\ \bibinfo {author} {\bibfnamefont {Paul-Gerhard}\
  \bibnamefont {Reinhard}},\ }\bibfield  {title} {\enquote {\bibinfo {title}
  {Self-consistent mean-field models for nuclear structure},}\ }\href {\doibase
  10.1103/RevModPhys.75.121} {\bibfield  {journal} {\bibinfo  {journal} {Rev.
  Mod. Phys.}\ }\textbf {\bibinfo {volume} {75}},\ \bibinfo {pages} {121--180}
  (\bibinfo {year} {2003})}\BibitemShut {NoStop}%
\bibitem [{\citenamefont {Nik{\v s}i{\'c}}\ \emph {et~al.}(2011)\citenamefont
  {Nik{\v s}i{\'c}}, \citenamefont {Vretenar},\ and\ \citenamefont
  {Ring}}]{niksic2011}%
  \BibitemOpen
  \bibfield  {author} {\bibinfo {author} {\bibfnamefont {T.}~\bibnamefont
  {Nik{\v s}i{\'c}}}, \bibinfo {author} {\bibfnamefont {D.}~\bibnamefont
  {Vretenar}}, \ and\ \bibinfo {author} {\bibfnamefont {P.}~\bibnamefont
  {Ring}},\ }\bibfield  {title} {\enquote {\bibinfo {title} {Relativistic
  nuclear energy density functionals: Mean-field and beyond},}\ }\href
  {\doibase 10.1016/j.ppnp.2011.01.055} {\bibfield  {journal} {\bibinfo
  {journal} {Prog. Part. Nucl. Phys.}\ }\textbf {\bibinfo {volume} {66}},\
  \bibinfo {pages} {519 -- 548} (\bibinfo {year} {2011})}\BibitemShut {NoStop}%
\bibitem [{\citenamefont {Dickhoff}\ and\ \citenamefont
  {Barbieri}(2004)}]{dickhoff2004}%
  \BibitemOpen
  \bibfield  {author} {\bibinfo {author} {\bibfnamefont {W.H.}\ \bibnamefont
  {Dickhoff}}\ and\ \bibinfo {author} {\bibfnamefont {C.}~\bibnamefont
  {Barbieri}},\ }\bibfield  {title} {\enquote {\bibinfo {title}
  {Self-consistent green's function method for nuclei and nuclear matter},}\
  }\href {\doibase 10.1016/j.ppnp.2004.02.038} {\bibfield  {journal} {\bibinfo
  {journal} {Prog. Part. Nucl. Phys.}\ }\textbf {\bibinfo {volume} {52}},\
  \bibinfo {pages} {377 -- 496} (\bibinfo {year} {2004})}\BibitemShut {NoStop}%
\bibitem [{\citenamefont {Som\`a}\ \emph {et~al.}(2013)\citenamefont {Som\`a},
  \citenamefont {Barbieri},\ and\ \citenamefont {Duguet}}]{soma2013}%
  \BibitemOpen
  \bibfield  {author} {\bibinfo {author} {\bibfnamefont {V.}~\bibnamefont
  {Som\`a}}, \bibinfo {author} {\bibfnamefont {C.}~\bibnamefont {Barbieri}}, \
  and\ \bibinfo {author} {\bibfnamefont {T.}~\bibnamefont {Duguet}},\
  }\bibfield  {title} {\enquote {\bibinfo {title} {\textit{Ab initio}
  gorkov-green's function calculations of open-shell nuclei},}\ }\href
  {\doibase 10.1103/PhysRevC.87.011303} {\bibfield  {journal} {\bibinfo
  {journal} {Phys. Rev. C}\ }\textbf {\bibinfo {volume} {87}},\ \bibinfo
  {pages} {011303} (\bibinfo {year} {2013})}\BibitemShut {NoStop}%
\bibitem [{\citenamefont {Tichai}\ \emph {et~al.}(2020)\citenamefont {Tichai},
  \citenamefont {Roth},\ and\ \citenamefont {Duguet}}]{tichai2020}%
  \BibitemOpen
  \bibfield  {author} {\bibinfo {author} {\bibfnamefont {Alexander}\
  \bibnamefont {Tichai}}, \bibinfo {author} {\bibfnamefont {Robert}\
  \bibnamefont {Roth}}, \ and\ \bibinfo {author} {\bibfnamefont {Thomas}\
  \bibnamefont {Duguet}},\ }\bibfield  {title} {\enquote {\bibinfo {title}
  {Many-body perturbation theories for finite nuclei},}\ }\href {\doibase
  10.3389/fphy.2020.00164} {\bibfield  {journal} {\bibinfo  {journal}
  {Frontiers in Physics}\ }\textbf {\bibinfo {volume} {8}} (\bibinfo {year}
  {2020}),\ 10.3389/fphy.2020.00164}\BibitemShut {NoStop}%
\bibitem [{\citenamefont {Chen}\ \emph {et~al.}(2017)\citenamefont {Chen},
  \citenamefont {Kaiser}, \citenamefont {Mei{\ss}ner},\ and\ \citenamefont
  {Meng}}]{chen2017}%
  \BibitemOpen
  \bibfield  {author} {\bibinfo {author} {\bibfnamefont {Q.~B.}\ \bibnamefont
  {Chen}}, \bibinfo {author} {\bibfnamefont {N.}~\bibnamefont {Kaiser}},
  \bibinfo {author} {\bibfnamefont {Ulf-G.}\ \bibnamefont {Mei{\ss}ner}}, \
  and\ \bibinfo {author} {\bibfnamefont {J.}~\bibnamefont {Meng}},\ }\bibfield
  {title} {\enquote {\bibinfo {title} {Effective field theory for triaxially
  deformed nuclei},}\ }\href {\doibase 10.1140/epja/i2017-12404-5} {\bibfield
  {journal} {\bibinfo  {journal} {The European Physical Journal A}\ }\textbf
  {\bibinfo {volume} {53}},\ \bibinfo {pages} {204} (\bibinfo {year}
  {2017})}\BibitemShut {NoStop}%
\bibitem [{\citenamefont {Papenbrock}\ and\ \citenamefont
  {Weidenm\"uller}(2020)}]{papenbrock2020}%
  \BibitemOpen
  \bibfield  {author} {\bibinfo {author} {\bibfnamefont {T.}~\bibnamefont
  {Papenbrock}}\ and\ \bibinfo {author} {\bibfnamefont {H.~A.}\ \bibnamefont
  {Weidenm\"uller}},\ }\bibfield  {title} {\enquote {\bibinfo {title}
  {Effective field theory for deformed odd-mass nuclei},}\ }\href {\doibase
  10.1103/PhysRevC.102.044324} {\bibfield  {journal} {\bibinfo  {journal}
  {Phys. Rev. C}\ }\textbf {\bibinfo {volume} {102}},\ \bibinfo {pages}
  {044324} (\bibinfo {year} {2020})}\BibitemShut {NoStop}%
\bibitem [{\citenamefont {Alnamlah}\ \emph {et~al.}(2021)\citenamefont
  {Alnamlah}, \citenamefont {Coello~P\'erez},\ and\ \citenamefont
  {Phillips}}]{alnamlah2021}%
  \BibitemOpen
  \bibfield  {author} {\bibinfo {author} {\bibfnamefont {I.~K.}\ \bibnamefont
  {Alnamlah}}, \bibinfo {author} {\bibfnamefont {E.~A.}\ \bibnamefont
  {Coello~P\'erez}}, \ and\ \bibinfo {author} {\bibfnamefont {D.~R.}\
  \bibnamefont {Phillips}},\ }\bibfield  {title} {\enquote {\bibinfo {title}
  {Effective field theory approach to rotational bands in odd-mass nuclei},}\
  }\href {\doibase 10.1103/PhysRevC.104.064311} {\bibfield  {journal} {\bibinfo
   {journal} {Phys. Rev. C}\ }\textbf {\bibinfo {volume} {104}},\ \bibinfo
  {pages} {064311} (\bibinfo {year} {2021})}\BibitemShut {NoStop}%
\end{thebibliography}

%

\end{document}